\documentclass[10pt,journal,compsoc]{IEEEtran}
\usepackage{paralist, tabularx}
\usepackage{graphicx}
\usepackage{wrapfig}
\usepackage{hyperref}

\usepackage[dvipsnames]{xcolor}

\newcommand{\visamount}{1,859}
\newcommand{\patternamount}{1,748}

\ifCLASSOPTIONcompsoc
  \usepackage[nocompress]{cite}
\else
  \usepackage{cite}
\fi
\begin{document}
\title{Revisiting the Design Patterns of\\Composite Visualizations}

\author{Dazhen Deng, Weiwei Cui, Xiyu Meng, Mengye Xu, Yu Liao, Haidong Zhang, Yingcai Wu% <-this % stops a space
\IEEEcompsocitemizethanks{\IEEEcompsocthanksitem D. Deng, X. Meng, M. Xu, Y. Liao, and Y. Wu were with the State Key Lab of CAD\&CG, Zhejiang University, Hangzhou, China, 310000. 
E-mail: \{dengdazhen, mengxiyu, mengyexu, yuliao, ycwu\}@zju.edu.cn. This work was conducted when Dazhen Deng was an intern at Microsoft Research Asia.
\IEEEcompsocthanksitem W. Cui and H. Zhang were with Microsoft Research Asia, Beijing, China, 100000. 
E-mail: \{weiweicu, haizhang\}@microsoft.com
\IEEEcompsocthanksitem Yingcai Wu and Weiwei Cui are the co-corresponding authors.
}% <-this % stops an unwanted space
\thanks{Manuscript received April 19, 2005; revised August 26, 2015.}
}


\markboth{Journal of \LaTeX\ Class Files,~Vol.~14, No.~8, August~2015}%
{Shell \MakeLowercase{\textit{et al.}}: Bare Demo of IEEEtran.cls for Computer Society Journals}



\IEEEtitleabstractindextext{%
%% Abstract section.
\begin{abstract}
Composite visualization is a popular design strategy that represents complex datasets by integrating multiple visualizations in a meaningful and aesthetic layout, such as juxtaposition, overlay, and nesting. With this strategy, numerous novel designs have been proposed in visualization publications to accomplish various visual analytic tasks. 
% These well-crafted composite visualizations have formed a valuable collection for designers and researchers to address real-world problems and inspire new research topics and designs. 
However, there is a lack of understanding of design patterns of composite visualization, thus failing to provide holistic design space and concrete examples for practical use. In this paper, we opted to revisit the composite visualizations in IEEE VIS publications and answered what and how visualizations of different types are composed together. To achieve this, we first constructed a corpus of composite visualizations from the publications and analyzed common practices, such as the pattern distributions and co-occurrence of visualization types. From the analysis, we obtained insights into different design patterns on the utilities and their potential pros and cons. Furthermore, we discussed usage scenarios of our taxonomy and corpus and how future research on visualization composition can be conducted on the basis of this study.
    
\end{abstract}

}

\maketitle

\IEEEdisplaynontitleabstractindextext

% \fi
\IEEEpeerreviewmaketitle

\section{Introduction}
% Para1: 
% what is visualization.
% common visualization widely used.
% data/task complexity increased ==> visual design evolved.
% although all new designs made, mostly still leveraging existing ones to composite.
% why? lower the learning curve, easy to generalize? 
% bring the definition of composite visualization
\IEEEPARstart{D}ata visualizations aim to visually represent data attributes to efficiently achieve the goals of analysis or storytelling~\cite{munzner2014visualization}.
For a long time, common visualizations (e.g., bar charts, line charts, and scatter plots) have been well-accepted by the public and widely adopted in business, education, and scientific research.
Because of the advancement in technology, complex data (e.g., large-scale, heterogeneous, hierarchical, and spatio-temporal) have become more and more available, and visualizations have also been evolving along with the complexity of analysis tasks, leading to the bloom of the visualization research community.
To address challenging analysis tasks, novel visual representations have been proposed from time to time, but the majority of research in this field still focuses on existing visualizations.
One common practice of leveraging existing visualizations for complex tasks is to compose different visual representations to exploit their advantages and make up for their disadvantages~\cite{javed2012exploring, keller2006matrices, laramee2004state, l2020comparative}.
In this work, we use the term \textbf{composite visualizations} to describe these visualizations.
Basically, they are a type of visualization that combines multiple visualizations in a meaningful and aesthetic layout~\cite{javed2012exploring}, such as juxtaposition, overlay, and nesting, to fulfill the need for specific data structures, analysis tasks, and usage scenarios.
Composite visualizations cover a large variety of design patterns.
%For the composite visualizations, the charts can be organized with different patterns to fulfill the need of different users and scenarios.
% For example, coordinated multiple views (CMV) is a well-established design pattern, which is flexible in layout and easy to understand, even for visualization novices.
% In a CMV, charts align side-by-side using a specific arrangement, such as sharing an axis or repeatedly listing same types of visualizations (e.g., \autoref{fig:juxtaposed_examples}).
% \ww{CMV just means juxtaposed? why not just call it juxtaposed?}
For example, a common design pattern is juxtaposing multiple charts side-by-side, which is flexible in layout and easy to understand, even for visualization novices.
To improve visual coherence, juxtaposed charts can be arranged in specific patterns, such as sharing an axis or repeatedly listing the same types of visualizations (e.g., \autoref{fig:juxtaposed_examples}).
% and coordinated by explicit visual hints (e.g., links and highlights) or implicit data flow controlled by user interactions (e.g., brushing and selecting).\highlight{can use the patterns we identify later to exemplify.}
In this way, a complex dataset can be visualized with multiple simple charts exhibiting different aspects of the data.
Because of the ease of implementation and understanding, the juxtaposition is widely used in visual analytics systems~\cite{chen2020composition}, fact sheets~\cite{wang2019datashot}, visual data stories~\cite{shi2020calliope}, etc.
Apart from juxtaposition, multiple visualizations are often compacted into a single view by overlaying (e.g., \autoref{fig:overlay_examples_1} \& \ref{fig:overlay_examples_2}) or nesting (e.g., \autoref{fig:nested_examples}).
By correlating the spatial and semantic relationships between graphical elements, such an integrated visualization is mainly tailored to reveal a specific type of pattern of the back-end data.
However, designing a successful composite visualization is not an easy task.
It requires not only an extensive knowledge base of visualization charts but also sufficient design skills to coherently present graphical elements for analysis tasks.
% ta is mainly  which requires extra considerations of graphical characteristics of these visualizations.
% is mainly tailored to address a specific analysis task.
% Such integration often requires extra considerations of the graphical characteristics of different visualizations and the spatial and semantic relationship between graphical elements, which helps reveal specific patterns of the back-end data.
%Because of their usefulness, numerous composite visualizations have been proposed in the visualization community.

On the other hand, there is a growing collection of visualization publications containing well-designed composite visualizations, which serve as a resource to reuse and inspire new research.
% For developers who design visual analytic systems to address real-world problems, design efficiency and feasibility are critical.
% Enumerating different combinations of visualizations will take a lot of time on engineering, but may not necessarily result in an effective design.
% Furthermore, it will be more cost-friendly to leverage concrete and validated design examples for further adaptation in a new task, instead of starting from scratch.
% For designers who design fancy visualizations for data communication, design efficiency and feasibility are critical.
To design a new visualization, enumerating different combinations of visualizations will take a lot of time to edit, but may not necessarily result in a design with promising visual effects.
It will be more cost-friendly to leverage concrete design examples for further adaptation to a new task, instead of starting from scratch.
From the perspective of research, a holistic design space of the composite visualizations might lead to new research topics (e.g., which are the most frequently used design patterns, and what are the reasons behind them?) and novel designs (e.g., is it possible to create efficient designs for specific tasks from the rarely occurring visualization combinations?).
%Such design patterns help to understand the real-world design practices for composite visualizations.
% To create novel designs, the researchers can learn from a holistic review of the design patterns and exploit the undiscovered area in the design space.
% Furthermore, to demonstrate the novelty of a new design, a holistic review of existing designs and comparison over the alternatives are necessary.
% First, it requires in-depth understanding of not only the analysis tasks but also visualization knowledge to select appropriate visual structures to encode the required data fields.
% , knowledge of visualization, and intelligence of aesthetic.
% In visualization publications, there is a large collection of composite visualizations, which serves as a resource for developers and researchers to transfer existing designs into new scenarios and carry out new designs. 

% Para 3: 具体化我们的问题和任务，希望得到的产出。
% 需要很明确的说出我们想做的事情:understanding real-world design practices for composite visualizations
% 然后，很明确的说我们主要要分析两点。为什么分析这两点而不是别的？分析这些有什么用。就要用上面的痛点背书呼应。
% 肯定不是practical guidelines，毕竟提供guideline不是我们要强调的
% 如果需要，也可以提到和过去工作的差别，说我们想做的事情以前都没做过
% However, we lack a comprehensive understanding of the current practices of composite visualizations in the community, thus resulting in the absence of practical guidelines.
In this work, we revisit the composite visualizations in VIS publications and try to understand their design practices from two perspectives.
First, from the perspective of visual components, \textit{what visualizations can be composed together as a composite visualization?}
Prior studies have explored composite visualizations in specific contexts (e.g., visual comparison~\cite{gleicher2017considerations, l2020comparative}), data (e.g., dynamic network~\cite{beck2014state}, multi-variate graph~\cite{nobre2019state}), or layout (e.g., juxtaposed views~\cite{chen2020composition}).
We opt to answer this question from a broadened scope of visualizations beyond specific tasks and data. 
Removing these restrictions, we can focus on visual designs and provide valuable exemplars for visualization development.
% , for example, the visualizations with specific layouts (e.g., CMV~\cite{chen2020composition}), of specific types (e.g., graph visualizations~\cite{} and glyph-based visualization~\cite{borgo2013glyph}), or for specific tasks (e.g., visual comparison~\cite{gleicher2017considerations, l2020comparative}).
Second, from the perspective of spatial relations, \textit{how can different visualizations be composed together?}
Prior studies~\cite{javed2012exploring, schottler2021visualizing, beck2014state} have summarized different design patterns for composite visualizations.
On the basis of these insights, we opted to revisit the patterns and conduct a quantitative analysis on how frequently different patterns are used~\cite{javed2012exploring}.
Such a demographic analysis will be helpful for spotting design trends, proposing design suggestions, and discovering the potential of under-explored design patterns.
Especially for researchers, a comprehensive survey can provide an overview of the community and inspire further research.
For example, a widely used design pattern might request further research of empirical studies to validate its efficiency.

In this work, we first construct a composite visualization corpus from IEEE VIS publications and decompose their designs into basic visualizations.
The decomposition enables us to identify composite visualizations and answer the first question.
Next, based on the decomposed visual designs, we formulated a two-level taxonomy of composition patterns to answer the second question.
In the taxonomy, the design patterns are identified according to the spatial relations and the semantic information conveyed.
To obtain an overview of the corpus, we revisited the visual designs based on the taxonomy and conducted statistical analysis on different design patterns.
% We first formulate a taxonomy for the design patterns of the composite visualizations based on the work of Javed and Elmqvist~\cite{javed2012exploring}.
% With the taxonomy, we construct a comprehensive visual design corpus from IEEE VIS.
% We then opt to systematically decompose and categorize the designs into our taxonomy.
% Second, we statistically analyze the design patterns of the composite visualizations across years and venues. 
% For each pattern, we then summarize design guidelines based on the data.
% Each guideline is accompanied with adequate real-world examples for the ease of understanding.
Finally, we construct an exploration system for the composite visualization corpus.
The system supports retrieving visualizations by type, composition pattern, and meta information.
The corpus, design pattern taxonomy and the exploration system can be viewed online: \url{https://composite-visualizations.github.io/}.
The contributions of this work include:
\begin{itemize}[$\diamond$]
    \item A taxonomy of composition pattern and a corpus with \patternamount{} composite visualization examples from IEEE VIS publications.
    \item An in-depth analysis of the statistics, utilities, advantages, and disadvantages of different composition patterns.
    \item Discussions of usage scenarios of the taxonomy and future research opportunities.
\end{itemize}
\section{Related Work}
This section introduces related studies about visualization configuration, design space of composite visualizations, and figure analysis to publications.

\subsection{Visualization Configuration}
Visualization configuration is a fundamental problem for visualization design and generation.
Previous studies on visualization configuration have extensively studied the composition of graphical elements for visualization rendering.
For example, Blackwell and Engelhardt~\cite{blackwell2002meta} termed the ``composition'' as the structure of graphical primitives (e.g., line and point).
Engelhardt~\cite{von2002language} further introduced a framework of syntax that recursively formulates a visualization to be a composite graphic object.
Based on the syntax, Engelhardt and Richard~\cite{engelhardt2018framework} investigated the ``DNA'' of visualization and proposed a grammar named VisDNA.
The study indicated various relationships between graphical primitives, such as grouping, nesting, and connecting, and scaling patterns of primitives, such as repeating.
Sedig and Parsons~\cite{sedig2016design} proposed a language that characterizes visual design patterns and the manners to fuse the patterns, such as self-similar nesting, layering, and juxtaposing.
Our work is similar to these ones in terms of leveraging the concept of ``composition'', but we focus on an entirely different granularity. Specifically, the composition patterns in these previous studies mainly focus on graphical primitives, such as line, point, and circle, which are more fundamental than the ones studied in this work.
For example, a bubble treemap is composed of self-nesting circles. 
In this work, the building blocks are high-level visualizations, which are the configurations of multiple graphical primitives. For example, a clustering heatmap is composed by stacking a heatmap and a clustering tree.

Visualization programming languages, such as ggplot (grammar of graphics~\cite{wilkinson2012grammar}) and Vega-Lite~\cite{satyanarayan2016vega}, also investigated how graphical primitives are composed during chart rendering.
In terms of composite visualizations, Vega-Lite supports fusing multiple charts by specifying the key of ``concat,'' ``layer,'' and ``facet'' in a declarative manner.
Nevertheless, Vega-Lite mainly focuses on the composition of charts with the Cartesian coordinate system.
More complex compositions such as nesting are not considered.
This study opts to revisit the design patterns of composite visualizations and provide insights into the design of grammar for more powerful visualization programming languages.

% % Reviewing composite visualizations can help us to gain an overview of how novel visual designs are configured in recent visualization research.

% Mackinlay~\cite{mackinlay1986automating} proposed a graphical presentation language, named APT, that enables the generation of composite charts, such as aligned bar charts.
% Our goal is to revisit the design patterns of composite visualizations in visualization research publications.
% and Vega-LiteThese languages also partially support the configuration of composite charts such as faceting and concatenating.

\subsection{Design Space of Composite Visualizations}
\begin{figure}[tb]
    \centering
    \includegraphics[width=\linewidth]{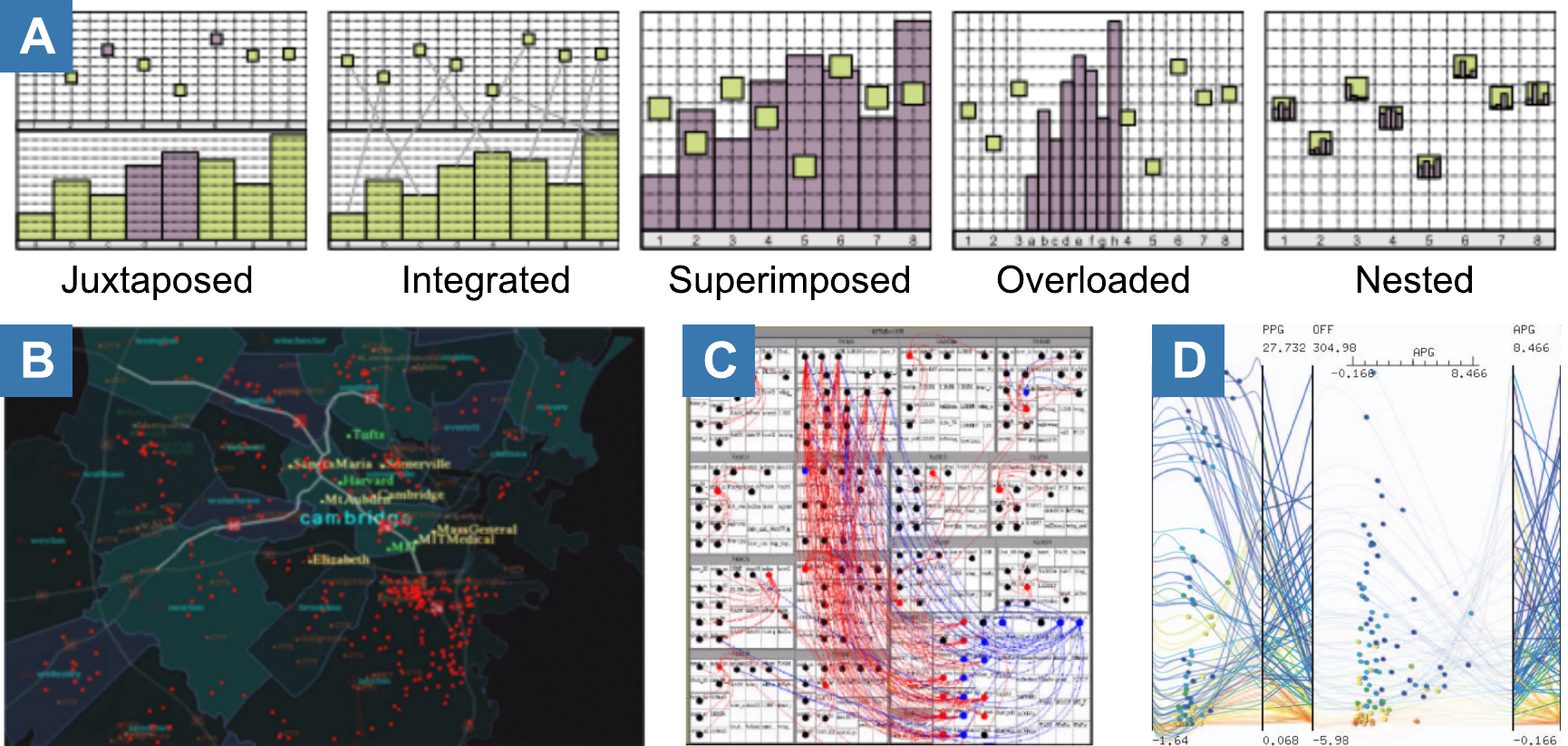}
    \caption{Taxonomy of composition patterns (A) proposed by Javed and Elmqvist~\cite{javed2012exploring} and examples~\cite{yuan2009scattering, fekete2003interactive, ishantha1995geospace} of superimposed views (B) and overloaded views (C \& D) provided in their paper.}
    % \vspace{-3mm}
    \label{fig:old_taxonomy}
\end{figure}
Javed and Elmqvist~\cite{javed2012exploring} proposed the term \textit{composite visualization view} as a theoretical model.
In their model, a composite visualization is described by its visual components, composition pattern, and data relationship.
Among them, the composition pattern, such as juxtaposition, integration, superimposition, overloading, and nesting, is used to describe how two visual components are spatially combined together (\autoref{fig:old_taxonomy}).
Inspired by their work, we use this model as a starting point and aim to revisit the design patterns of composite visualizations in IEEE VIS publications.
On top of that, our work can better identify composition patterns, obtain new insights about composite visualizations, and facilitate new usage scenarios.
%  compared to Javed and Elmqvist's work.
First, we have proposed a more refined taxonomy for revisiting the large corpus of visual designs in IEEE VIS.
For example, we differentiate the overloaded views (e.g., \autoref{fig:old_taxonomy}B) and superimposed views (e.g., \autoref{fig:old_taxonomy}C \& D) considering the use of coordinate systems in our taxonomy.
% The original taxonomy is not clear enough for classification when annotating a large corpus of visual designs.
% Javed and Elmqvist's taxonomy is crafted by inspecting tens of visual designs that combine multiple visualizations in the same view.
% However, when annotating a large corpus of visual designs from thousands of papers, some definitions in their taxonomy are not clear enough for classification.
% Therefore, we incorporated them into the same type (overlaid visualization) and identified additional sub-types with a visual-based approach.
% The newly proposed sub-types are feasible to categorize visualizations with different back-end data structures based on observable features (such as with or without shared coordinates).
Second, we obtained new insights into design patterns with quantitative analysis.
For example, a correlation analysis on visualization types within individual composition patterns presents a general usage preference of visualization combinations in the visualization community.
% With the large corpus, we gained knowledge about the overall statistics of different design patterns and their advantages, disadvantages, and general utilities from the samples.
Third, the taxonomy and corpus can facilitate new usage scenarios, such as helping visualization designers and researchers in improving the efficiency of survey and design with an exploration system.
% A detailed discussion is provided in \autoref{sec:secnario}.

Other studies related to visualization composition may target specific data~\cite{schottler2021visualizing, nobre2019state, beck2014state} or tasks~\cite{gleicher2011visual, l2020comparative}.
% Given specific data, the attribute type and data structure restrict the accessibility of visualization types, thus limiting the visualization combinations. 
Researchers have explored how to combine different visualizations to represent data with specific structures or types.
For example, for dynamic graphs, Beck et al.~\cite{beck2014state} studied how node-link diagrams and timelines are juxtaposed, superimposed, and integrated together to encode the temporal information.
For geospatial networks, Sch\"ottler et al.~\cite{schottler2021visualizing} analyzed the combination of node-link diagram and map based on the taxonomy of Javed and Elmqvist's~\cite{javed2012exploring}.
Nobre et al.~\cite{nobre2019state} studied the juxtaposed, integrated, and overloaded patterns of matrix and node-link diagrams in multivariate networks.
Another group of studies focuses on how different compositions of visualizations affect the efficiency of a specific task.
For example, Gleicher et al.~\cite{gleicher2011visual} proposed three composition patterns of visualizations for visual comparison, including juxtaposition, superimposition, and explicit encoding.
L'Yi et al.~\cite{l2020comparative} further reviewed and summarized visual comparisons under these patterns, and presented several practical design guidelines.
% Finally, from the perspective of layout, the design space of juxtaposed views (or coordinated multiple views~\cite{javed2012exploring}) has been well studied.
% % researchers have studied the design space of multiple-view visualizations, the composite visualizations with 
% Baldonado et al.~\cite{wang2000guidelines} contributed four rules (diversity, complementarity, parsimony, and decomposition) to organizing multiple-view visualizations.
% Qu et al.~\cite{qu2017keeping} emphasized the consistency between different views.
% In addition to normal computer screens, researchers also proposed principles and methods for adjusting the layout of multiple views in mobile devices~\cite{langner2017v, sadana2016designing}, large display~\cite{langner2018multiple}, and cross-platform interactions~\cite{horak2019vistribute}.

% To gain an overview of the practices of multiple-view visualizations, Chen et al.~\cite{chen2020composition} adopted a hierarchical coding scheme to decompose the layout of the multiple-view visualizations, and recommended by querying similar state-of-the-arts.
Compared to these studies, we removed the constraints of specific data or tasks, and analyzed the design patterns of composite visualizations with an extended scope.
Our study presents a corpus including diverse visualization types and layouts and an overview of the state-of-the-art composite visualizations.

% under different tasks~\cite{gleicher2011visual, gleicher2017considerations, l2020comparative}, visual representations~\cite{beck2014state, nobre2019state}, and styles~\cite{chen2020composition, hadlak2015survey}.
% However, the meaning of these composition patterns are not consistent.
% From the level of tasks, existing studies~\cite{gleicher2011visual,gleicher2017considerations,l2020comparative} focus on how different visual composition can help achieve specific tasks.
% For example, L'Yi et al.~\cite{l2020comparative} have studied the effectiveness of juxtaposition and superposition on the tasks of visual comparison.
% Besides, the data types are generally related to the visual representations.
% For example, the network data (e.g., dynamic network~\cite{beck2014state} and multi-variate network~\cite{nobre2019state}) can be visualized with different compositions of node-link diagram and matrix; and spatio-temporal data is commonly exhibited by 
% In addition, researchers have studied the design space of the CVVs with specific styles. 
% CMV, the CVVs with juxtaposed pattern~\cite{javed2012exploring}, is a typical case that different visualizations can be organized in a hierarchical structure~\cite{chen2020composition}.
% There are also studies using hybrid specifications.
% For instance, Hadlak et al.~\cite{hadlak2015survey} explored the design space of the graph visualizations that with multiple facets.

\subsection{Figure Analysis of Visualization Publications}
% Literature analysis is an important technique to understand the evolution and future trend of a research field.
% % Current methods mainly focus on the text, citation, authors, and other metadata of publications~\cite{federico2016survey}.
% % Based on these data, a series of visual analytics tools and techniques have been proposed, such as VisList~\cite{gorg2012combining, isenberg2016vispubdata}, CiteVis~\cite{stasko2013citevis}, CiteWiz~\cite{elmqvist2007citewiz}, and CiteRiver~\cite{heimerl2015citerivers}.
% In the visualization community, literature datasets have been proposed to reduce the burden of accessing clean and comprehensive data.
% Isenberg et al.~\cite{isenberg2016vispubdata} compiled the conference proceedings and contributed a comprehensive and constantly updating dataset, vispubdata.org, that contains essential information (e.g., title, abstract, author names, author keywords, and references) of all papers in IEEE VIS.
% In addition to the author keywords from publications, Isenberg et al.~\cite{isenberg2016visualization} proposed a novel keyword dataset through multi-pass manual coding and analyzed the topic changing of the visualization community.

% In recent years, we witness developing interest in analyzing the figures from visualization publications.
In addition to survey papers, researchers also analyzed visualization publication figures.
Li et al.~\cite{li2018toward} conducted a memorability study with SciVis figures.
Zeng et al.~\cite{zeng2020vistory} contributed VIStory, a technique for exploring figures in VIS publications. 
Chen et al.~\cite{chen2021vis30k} adopted object detection models to extract the figures and tables in IEEE VIS publications and proposed VIS30K.
These studies mainly focus on perception tasks and analytical techniques for figures, instead of the visual designs inside.

% Furthermore, researchers are interested in the charts inside the figures, where the research interests of this work also lie in.
For visual designs, some studies explore how different visualizations are distributed in the figures.
Deng et al.~\cite{deng2021visimages} collected figures from VIS publications and annotated the types and positions of visualizations.
However, they only considered the co-occurrence of visualizations in the figures and failed to answer how different visualizations are composed, which is more useful for designers and researchers.
Chen et al.~\cite{chen2020composition} explored the composition and configuration patterns of multiple-view visualizations (MV) consisting of juxtaposed views.
% They constructed a MV corpus by annotating the view types and positions in the user interface (UI) screenshots, and analyzed the types, alignments and shape of the views.
They discovered ``many novel designs with compound view types,'' which indicates more complex but under-explored design patterns other than juxtaposition.
Therefore, in this work, we moved a step further and studied the composite visualizations within single views, including the types of visual components and design patterns.
%Differently, we focused more on the relative positions between the composed visualizations and the semantic information behind the design patterns.
% However, they only considered the juxtaposed views and ignored more complex composition patterns, such as superimposition and nest~\cite{javed2012exploring}.

\section{Terminology}
\label{sec:terminology}

\begin{figure}[tb]
    \centering
    \includegraphics[width=0.9\linewidth]{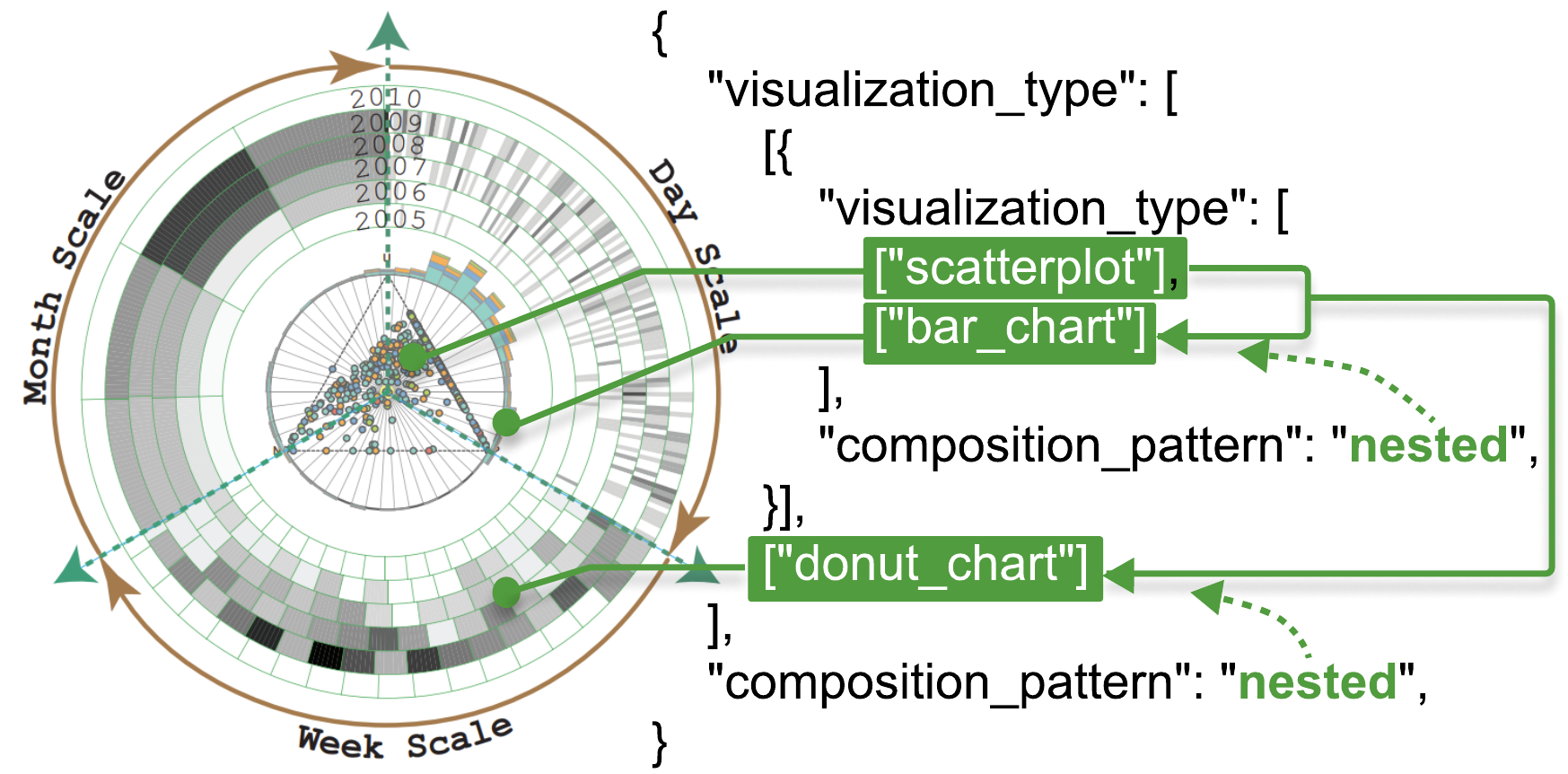}
    \caption{An example of composite visualization. OpinionSeer~\cite{wu2010opinionseer} is composed of scatterplots, bar charts, and donut charts. The composite visualization can be defined recursively with a hierarchical structure.}
    % \vspace*{-5mm}
    \label{fig:example}
\end{figure}

%Our terminology is built upon the theoretical model of composite visualization views proposed by Javed and Elmqvist~\cite{javed2012exploring}.
%In this work, a composite visualization is composed of basic visualization types with composition patterns.
A composite visualization is defined by basic visualizations and composition patterns.

\begin{itemize}[$\diamond$]
    \item \label{sec:basic-types}\textbf{Basic visualizations} are the components (or building blocks) of composite visualizations, referring to different types of visual representations, such as bar chart, parallel coordinate plot, and map. 
    It is noted that a basic visualization can be broken into smaller elements, such as marks, axes, and legends, but in composite visualizations the smallest building block is basic visualization.\\
    A series of studies attempt to classify visualizations~\cite{lohse1994classification, chi2000taxonomy, harris1999information, heer2010tour, meirelles2013design}.
    For example, Harris~\cite{harris1999information} presented an exhaustive categorization of visualization types and indexed them by alphabetical order.
    Meirelles et al.~\cite{meirelles2013design} categorized visualizations by data structures.
    Heer et al.~\cite{heer2010tour} classified the charts by their data structures and tasks.
    We choose Borkin's taxonomy~\cite{borkin2013makes} as the visualization type taxonomy because it covers most of the aforementioned taxonomies and contains additional up-to-date visualization types.
    As demonstrated by a previous study~\cite{deng2021visimages}, this taxonomy can serve as a useful vocabulary for researchers to classify the visualizations in visualization research publications.
    % analyze the design patterns of the visual designs, such as data stories~\cite{shu2020makes, shi2020calliope} and multiple-view visualizations~\cite{chen2020composition, wang2019datashot}.
    % is the most comprehensive one that summarizes previous taxonomies and includes up-to-date visualization types (e.g., Sankey diagram and chord diagram), which is suitable for annotating the visual designs in the visualization research papers.
    Borkin's taxonomy classifies visualizations according to data structures, visual encodings, and tasks. The taxonomy has two classification levels (12 first-level categories, each with several second-level sub-types).\\
    % In this work, we adopted the second level.\\
    We identified three issues when using Borkin's taxonomy for the goal of this work.
    First, some types have multiple names and definitions, such as histograms and bar charts.
    We unify these types for simplicity.
    Second, some types are semantically similar, such as graphs and trees. The tree is a special case of the graph that has a hierarchical data structure. The semantic similarities provide more fine-grained differentiation between classes that help us gain more insights into the composition patterns. For example, graphs are commonly overlaid on map visualizations but trees are not.
    To avoid duplicated annotations, we only assign one class to each visual component based on its shapes. For example, a visual component is assigned one of the labels ``tree'' or ``graph''. We will choose the label that more accurately describes the visual components.
    Third, Borkin's taxonomy does not cover scientific-specific visualizations (e.g., volume rendering), so we add a type named ``SciVis''.

    \item \textbf{Composition patterns} refer to the visual relationships between components in a composite visualization. In this work, we propose a taxonomy of composition patterns considering relative positions (e.g., overlapping) and attribute relationships (e.g., type and style) between components. According to the terminology in Javed and Elmqvist~\cite{javed2012exploring}, where composition patterns describe the usages of space and a relation between the visual components in composite visualization. Therefore, we regard composition patterns as a set of reusable configurations for the generation of a new visualization design given basic visualizations, which demonstrates general visualization design patterns.
    % \item Composite visualization is a visualization that composed of multiple basic visualization types.
    % \item \textit{Instance relation} illustrates the potential numbers of the composed visual structures. The counting relation can be categorized into binary relations (one-to-one, one-to-many, and many-to-many) and unary relations (many-as-one).
\end{itemize}

Given these two characteristics, we define a \textbf{composite visualization} by its components organized with composition patterns, where a component can be either a basic visualization or another composite visualization.
Using this recursive definition, a composite visualization can be represented using a hierarchical structure (\autoref{fig:example}).
Note that for a composite visualization, composition patterns are necessary, which is different from other multiple-view visualizations (or user interfaces) that can be loosely defined as a group of charts placed together~\cite{chen2020composition, roberts2007state}.

\section{Corpus Construction}
% The construction of composite visualization views (CVVs) corpus is motivated by Javed and Elmqvist~\cite{javed2012exploring}, who have summarized five design patterns of CVV through literature review.
This section demonstrates how we constructed the corpus of composite visualizations based on VisImages~\cite{deng2021visimages}.
% To construct a corpus of composite visualizations, we first collected visualizations from the figures of visualization publications based on a former dataset, VisImages~\cite{deng2021visimages}.
% Second, we analyzed and labeled the composition patterns in the collected examples.

\subsection{Collecting Figures and Designs}
\label{sec:anno_design}
Our main goal is analyzing composite visualizations in research publications of visualizations.
Therefore, we constructed the corpus based on VisImages, a dataset that collects figures (as well as the basic visualization types and positions in the figures) from IEEE VIS proceedings. 
% IEEE VIS is a top venue that contains the state-of-the-art visualization designs.
We focused on the papers from 2006 (when VAST was established) to 2020 and obtained 19,910 figures from 1,963 papers.
Many other visualization venues, such as EuroVis, ACM CHI, Diagrams, and Infovis journal, also contain high-quality visualization designs. As a starting point, we primarily focus on IEEE VIS and leave the analysis of these venues to future research.
% \deleted{We started by collecting the papers of IEEE VIS (InfoVis, VAST, and SciVis), the top venues that contain the state-of-the-art visualization designs, from 2006 (when VAST was established) to 2020.
% The pdf files of the papers were collected from the conference proceedings with the help of vispubdata.org\footnote{https://sites.google.com/site/vispubdata/home}.}

% \deleted{From the papers, we collected figures with a method similar to the ones used by Chen et al.~\cite{chen2020composition} and Lee et al.~\cite{lee2017viziometrics}.
% Specifically, we used PDFFigures 2.0~\cite{clark2016pdffigures} to extract figures and corresponding captions.
% The figure positions were also extracted and represented by a rectangular bounding box.
% To ensure the data quality, we checked and revised the figure boundaries and captions through crowdsourcing.
% In total, we obtained a corpus consisting of 1,963 papers and 19,910 figures.}

However, the collected figures have purposes that are not suitable for follow-up composite visualization analysis.
For example, a large part of the figures is statistical figures used in evaluations, which should be excluded from the corpus.
Therefore, we only kept the figures containing original visualization designs.
We established three criteria for the filtering.
% Therefore, we exclude the figures for illustrating the concept and process (e.g., flow chart, photo, and description) and reporting experiment results.
First, we only selected figures containing visualization designs used for data analysis.
We excluded figures that illustrate models, frameworks, experiment results, etc.
We also excluded figures showing visualization designs from previous papers, such as figures in survey papers~\cite{johansson2015evaluation}.
Second, if the components of a design appear in several figures, we selected the one with the most components to maximize the design integrity.
%Second, we selected the figures with more views if a visualization was included in multiple figures. 
%For example, there were figures for system interface and detailed visual designs of each view.
%In this case, we selected the figures rendering the system interface.
Finally, if there are multiple figures duplicated in terms of visualization design, we selected the first one.

% When selecting figures, we also specified the position of designs of interest with rectangular bounding boxes, because there may be several subfigures and visual designs in a figure.
% The drawing of bounding boxes ensured each design only appears once.

We developed an interactive tool for figure selection and design annotation.
The tool helps users verify if the figures meet the three criteria and locate the visual designs in the verified figures.
%After locating visualization designs with bounding boxes, we also used the tool to specify the type of visualizations from ``multiple-view visualization'' or ``single-view visualization''.
Using this tool, three authors independently filter the corpus based on the three criteria.
% During the annotation process, captions to verify if the figures meet the criteria.
% Another author double-checked the figures and recorded the ones he/she disagreed.
The inconsistent results are discussed and resolved using the majority voting rule.
As a result, we filtered 1,353 figures from the 19,910 figures and collected 1,565 visualization designs from the filtered figures.
%Among them, there are 727 (65.7\%) multiple-view visualizations and 379 (34.3\%) single-view visualizations.

\subsection{Annotating Composite Visualizations}
For each composite visualization, we attempted to annotate the composition patterns for further analysis.
In VisImages, all basic visualization types and their positions in the figures are identified using the type taxonomy proposed by Borkin et al.~\cite{borkin2013makes}. We further verified and revised the visualizations based on the taxonomy descriptions in \autoref{sec:basic-types} with the interactive tool.
Thereafter, the visualization designs are decomposed into a series of basic visualization types.
Please note that in some cases, a visualization can be assigned with multiple type labels: when the visual component is a heatmap (defined based on color encodings) and other visualization types (based on shapes) simultaneously. 
However, the multiple label issue is not prevalent (19/1,859). We retain the type heatmap as it can provide information about how composition pattern is used for this particular visualization type.
Based on the decomposition, we annotate the composition patterns in a bottom-up manner.
First, we analyzed and collected different composition patterns according to the spatial and attribute relationships in each example, then built a taxonomy of composition patterns based on the collected patterns.
Second, we labeled all designs in the corpus with the taxonomy and filtered the composite visualizations for the follow-up statistical analysis.

\subsubsection{Analyzing Composition Patterns}
\begin{figure*}
  \centering
  \includegraphics[width=\textwidth]{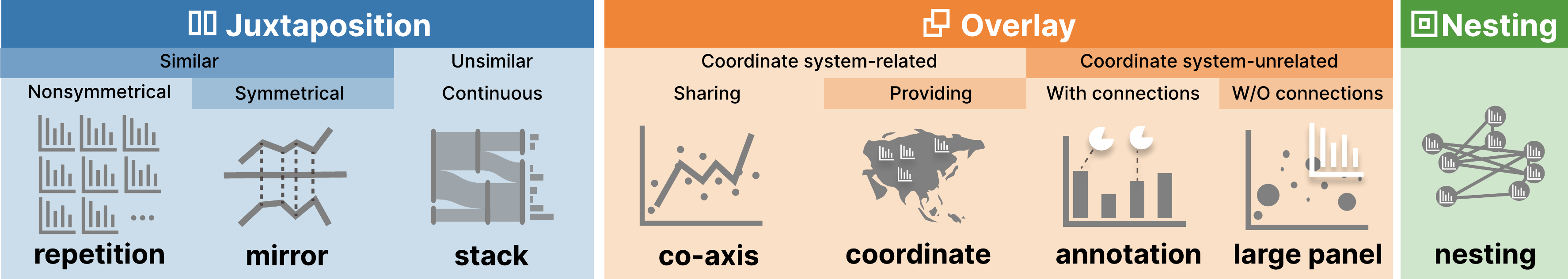}
  \caption{A two-level taxonomy of Composition Patterns. Composite visualizations are classified according to overlapping relations at the first level, and further classified into sub-types according to geometric (symmetric or continuous) and semantic relation (e.g., sharing coordinate systems or providing coordinate systems for other components). }
  \label{fig:teaser}
\end{figure*}
Cross-referring the taxonomy proposed by Javed and Elmqvist~\cite{javed2012exploring} and the collected corpus, we constructed a two-level classification of composition patterns.
We term the classification as a taxonomy because we follow a series of rules to exhaustively and exclusively divide the design space into several sub-spaces.
At the top level, according to the overlapping relationships between basic visualizations, we identified three patterns, namely, \textbf{juxtaposition}, \textbf{overlay}, and \textbf{nesting}.
Compared to the taxonomy proposed by Javed and Elmqvist (\autoref{fig:old_taxonomy}), ours has several major changes.
First, the integrated views are considered as juxtaposition visualizations in our taxonomy, because we consider the explicit visual links and underlying data flow as coordination methods between juxtaposed visualizations.
Second, the superimposed and overloaded views are merged as overlay visualizations, because their views are both composed by visually overlaying visualizations on others.
% \begin{table*}[!htb]
%     \centering
%     \caption{Detailed categorization of the composition patterns.}
%     \begin{tabular}{|c|c|c|c|c|c|c|c|} 
%     \hline
%     \multicolumn{3}{|c|}{\textbf{Juxtaposed}}& \multicolumn{4}{c|}{\textbf{Overlaid}}& \textbf{Nested}  \\ \hline
%     \multirow{2}{*}{similar}& symmetric  & asymmetric  & equal  & \multicolumn{3}{c|}{\textbf{accompanied}}  & \multirow{4}{*}{\textbf{nested}}  \\ \cline{2-3}\cline{4-7}
%     &\textbf{mirrored}& \textbf{repeated} & \multirow{3}{*}{unequal} & spatially-coordinated&\multicolumn{2}{c|}{non-spatially-coordinated} &\\ \cline{1-3}\cline{5-7}
%     \multirow{2}{*}{dissimilar} & axis-sharing & non-axis-sharing &  & \multirow{2}{*}{\textbf{coordinated}} & whole & \textbf{auxiliary}&\\\cline{2-3}\cline{6-7}
%     & \textbf{stacked} & \textbf{co-occurring} &&& part  & \textbf{annotated}&\\\hline
%     \end{tabular}
%     \label{tab:taxonomy}
% \end{table*}
More importantly, for each composition pattern, we further identified several sub-types and contributed a second-level taxonomy based on more fine-grained visual features.
The proposed taxonomy and the features to identify each type are presented in \autoref{fig:teaser}.
For juxtaposition patterns, we referred to the Gestalt principle and firstly identified visualizations with components of the same visual structures or different visual structures following the similarity rule.
For the ones composed of similar visual components, we further identified \textbf{repetition patterns} and \textbf{mirror patterns} following the symmetry rule, while for the ones with different structures, we identified \textbf{stack patterns} based on the continuity rule.
% For the juxtaposed charts occurring in the same visual design, we consider that they are coordinated to facilitate visual analytic through specific relations.
% If not belonging to one of the repeated, mirrored, and stacked, the relations will be categorized into an \textbf{others} type.
For overlaid patterns, we first identified visualizations whose components have coordinate relation and that have no coordinate relation.
For the ones whose components have coordinate relation, we identified \textbf{co-axis patterns} in which different components share the same coordinate system and \textbf{coordinate patterns} in which one component serves as a coordinate system for other components.
For the ones whose components do not have coordinate relation, we differentiate \textbf{annotation patterns} and \textbf{large-panel patterns} based on the existence of connections between the child components and parent components.
% also identified four sub-types, including \textbf{accompany}, \textbf{coordination}, \textbf{annotation}, and \textbf{large-panel}.
% These four types are identified based on the sizes, the uses of coordinates, and graphical relationships.
% The detailed categorization of the composition patterns is shown in Table~\ref{tab:taxonomy}.

\subsubsection{Annotating Composition Patterns}
With the basic visualizations and composition pattern taxonomy, we further aggregated composition patterns based on different basic visualization types.
To ensure the completeness of the analysis, we analyzed the composition patterns by enumerating all possible combinations of the types. For example, imagine a visualization that multiple glyphs of bar charts are distributed on a map, and the map is a heatmap at the same time. We will assign three basic visualization types to the visualization and obtain three visualization type pairs.
For each pair of types in a composite visualization, we annotated its composition pattern.
%We carefully distinguish the elements and relationships that are not belonging to the designs themselves but added by original authors during figure creation for illustration purposes.
Since a composite visualization may have three or more components and is defined recursively, the annotation is also performed in a bottom-up and recursive manner.
Similar to basic visualization types, all composition patterns were annotated and verified by at least two authors.
All inconsistencies are resolved by involving a third author and the majority voting rule.
In total, we obtained \patternamount{} composition patterns from \visamount{} visualization type pairs.
% The labelling relied on an interactive tool (Fig.~\ref{fig:interface}) with functions to select the basic visualizations from a visualization list and specify a composition pattern between them.
% Once specifying a composition pattern, a composite visualization is discovered and added to the visualization list.
% The newly discovered composite visualization can be further selected as a component of more complex visualizations.
% The composition patterns can be divided into two classes according to the equality.
% The equal relations include mirrored, repeated, stacked, and accompanied views.
% The rest types are unequal relations, in which there are child view and parent view.
% For the equal relations, we selected all visualizations from the visualization list.
% For the unequal relations, we selected the visualizations belonging to the child view and parent view separately.
% After the view specification, we chose the relation types.
% For the composite visualizations with more than one layers, we firstly specify the child nodes.
% The specified child nodes can be selected and operated as the individual visualizations to form more complex visualizations.

% After the process of decomposition and aggregation, we obtained xx composite visualizations from xxx designs.

% \subsubsection{Rule of statistics}
Please note that, in a composite visualization, the combination of two visualization types is only counted once to avoid redundancy, regardless of the instance numbers of each type.
For example, in a scatterplot matrix, we only count the combination of \textit{scatterplot} + \textit{matrix} once.
% We adopt this rule of statistics because it can avoid counting a rare pattern for many times because of the frequent use in a single visual design.
% For example, the bar chart coordinated in the parallel coordinates plots in Fig.~\highlight{[]} will count eight times and becomes a frequent pattern without this rule.
% If a composite visualization has more than one layers and the components at specific layer
% This rule applies on each layer of the composite visualizations.
% Second, sometimes it is unable to count the actual number of occurrence due to the small sizes and occlusion, as shown in Fig.~\highlight{[]}.  
\section{Composite Visualizations in IEEE VIS}
\label{sec:design_patterns}

\begin{figure}[tb]
    \centering
    \includegraphics[width=\linewidth]{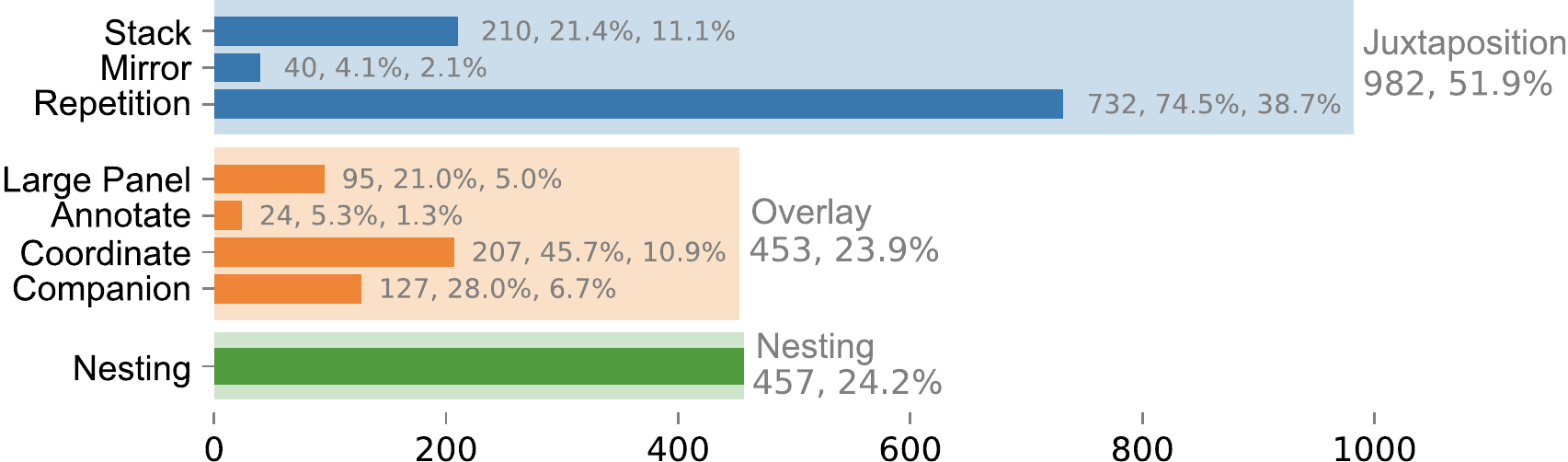}
    \caption{Distribution of composition patterns. 
    % Distributions of design patterns at different levels are visualized with a hierarchical bar chart. For each sub-type, we present the total number, proportion in the parent type, and proportion in the corpus.
    }
    \label{fig:overview}
\end{figure}

\autoref{fig:overview} shows the overall statistics of composition patterns.
In this section, we present details for each pattern by reporting the numbers and summarizing their utility.
% In this section, we introduce the statistics of different design patterns.
% The overall distributions of the design patterns are shown in \autoref{fig:overview}.
% In addition, we use matrices to visualize the distribution of 
% 11
%  type combinations of different design patterns (denoted as combination matrices).
% \highlight{Order: definition, statistics, scenario, advantages, disadvantages, design suggestions.}

\subsection{Juxtaposition}
For juxtaposition visualizations, their components do no overlap and are positioned side by side.
% The juxtaposition visualizations here are different from the coordinated multiple views (or multiple-view visualizations) on the semantic information.
% The juxtaposed views also refer to coordinated multiple views (or multiple-view visualizations).
Although a user interface (UI) may also consist of visualizations without overlapping, it is not considered a juxtaposition visualization in the context of this study.
The component interrelationships in a UI are considerably looser than those in a visualization with juxtaposition patterns.
As a rule of thumb, we consider a UI an arbitrary placement of visualizations or interface components (e.g., buttons, sliders, and progress bars) and only extract the visualizations.
% However, juxtaposition visualizations do not consider user interactions but concern charts bound with data and meaningful layouts between the charts.
% ...\ww{explain and emphasize the difference between this and ui.}

Compared to overlay and nesting where visual components are overlaid on or contained by other components, juxtaposition offers a flexible and clear layout for visual components without visual occlusions.
Juxtaposed components interrelate through visual links or data flow.
Visualization with juxtaposition patterns take up 53.8\% (941/\patternamount) of all composition visualizations, which is the most frequent pattern.
We identified three sub-patterns, namely, \textit{repetition}, \textit{mirror}, and \textit{stack}, corresponding to the similarity, symmetry, and continuity rules of the Gestalt principle.
% However, in juxtaposition visualizations, the separated visualizations may disperse the focus of the users when the number of visualizations increases.
% Therefore, to guide the users, it would be better to use specific design patterns to improve the visual salience of the views.

\subsubsection{Repetition Patterns}
\setlength\intextsep{0pt}
\setlength{\columnsep}{4pt}
\begin{wrapfigure}{l}{0.05\textwidth}
    \includegraphics[width=0.048\textwidth]{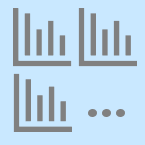}
\end{wrapfigure}
\textbf{Definition}: repetition refers to the juxtaposition visualizations which are of the same structure (visualization type or composite visualization), but their components are not symmetrical with respect to coordinate axes.
Repetition patterns are the most common (41.8\%, 730/\patternamount) among all patterns.
% From the samples, we discovered that repetition visualizations are more frequently used in system interfaces (79.2\%) than single-view visualizations.
% \begin{figure*}[!hbt]
%     \centering
%     \includegraphics[width=\textwidth]{figures/juxtaposed_examples.pdf}
%     \caption{Distribution of the number of repeated charts in a visualization.}
%     \label{fig:juxtaposed_examples}
% \end{figure*}

\begin{figure*}[!htb]
    \centering
    \includegraphics[width=\textwidth]{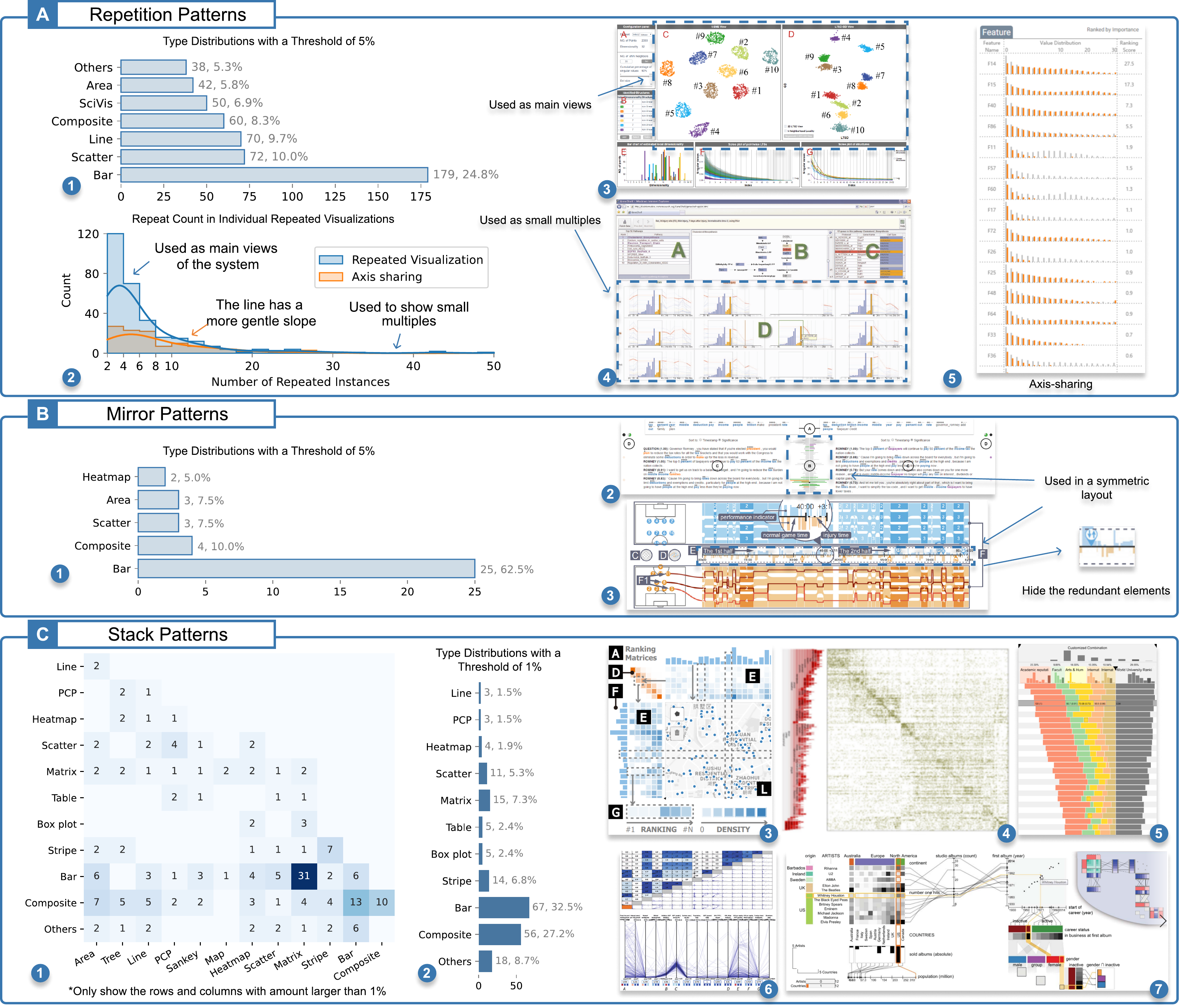}
    \caption{Statistics and examples of repetition visualizations (A: A3~\cite{xia2017ldsscanner}, A4~\cite{kim2009geneshelf} \& A5~\cite{liu2017visual}), mirror visualizations (B: B2~\cite{el2018topic} \& B3~\cite{wu2018forvizor}), and stack visualizations (C: C3~\cite{weng2018srvis}, C4~\cite{bilal2018conv}, C5~\cite{gratzl2013lineup}, C6~\cite{johansson2009interactive}, \& C7~\cite{gratzl2014domino}).}
    \label{fig:juxtaposed_examples}
    % \vspace{-3mm}
\end{figure*}

\autoref{fig:juxtaposed_examples}-A1 shows the occurrences of visualization types with repetition patterns.
We can see that the distribution is highly skewed.
Bar charts (25.1\%, 179), scatterplots (10.0\%, 72), and line charts (9.7\%, 70) are the most frequently used. Besides, composite visualizations (8.3\%, 60) and SciVis (6.9\%, 50) are also popular with repetition patterns.

\autoref{fig:juxtaposed_examples}-A2 shows the distribution of repetitions in individual repetition visualizations.
The long tail distribution indicates two different usage scenarios.
On the one hand, some visualizations may contain a large number of components (up to 100 in our collected examples), as shown in \autoref{fig:juxtaposed_examples}-A4 \&~\ref{fig:juxtaposed_examples}-A5.
These examples generally target lightweight usage, such as browse, selection, or overview, referring to small multiples in other research~\cite{van2013small}.
The information presented in a repetitive unit is often simple for better readability, and the components are commonly shown in a list view with a scroll bar to handle the scalability.
% We further investigated the concrete samples on their numbers of repeated components to understand the usage of repetition visualizations. 
% As shown in \autoref{fig:juxtaposed_examples}, the numbers of repeated components have a long tail distribution, indicating two different usage scenarios.
% First, from the tail of the distribution, we discovered that repetition visualizations can support a large number of components (up to 100).
% The repetition visualizations in this case (\autoref{fig:juxtaposed_examples}~A4) are used to show a large number of data series for browse, selection, and further exploration, referring to small multiples in other research~\cite{van2013small}.
% The information revealed in each component is relatively simple, showing the overview of data series.
% The components are commonly shown in a list with scroll bar to handle the scalability. 
On the other hand, from the peak of the distribution, where the repetition counts lie in $[2,4)$, we discovered a different usage scenario of exploration.
%In addition to small multiples, the repetition visualizations can also serve as main views in the system.
For example, in \autoref{fig:juxtaposed_examples}-A3, two scatterplots are used to show data projection with t-SNE and LTSD-GD, respectively~\cite{xia2017ldsscanner}. 
This visualization is used as the main view for showing details of the data, requiring a relatively large space.
%Therefore, the numbers of components mostly lie in $[2,4)$ , only showing the most important aspects of data.
Among the visualizations with repetition patterns, 38\% consist of components that share the same axis.
The repetition distribution in this subset also follows a long tail pattern, but with a more gentle slope.
% The gentle slope indicate that when the component number is small, many designs choose to visualize the axes of each component independently, because different components may show different data attributes with different scales.
With shared axes, repeated components can be easily compared at the same scale.

Repetition patterns have three main advantages. 
First, the visual similarity of repeated components provides a strong visual hint of grouping according to the Gestalt principle~\cite{rosenholtz2009intuitive}.
Second, repetition visualizations can help users in exploring and comparing multiple items, but might not be the best choice compared to mirrored or superimposed layouts~\cite{l2020comparative,ondov2018face}.
Third, repetition visualizations are easy to implement because codes can be reused conveniently.
However, repetition patterns also have disadvantages. 
First, it is difficult to compare when the number of components gets large because the distance between targeted components might be large~\cite{l2020comparative}.
Second, because of the same appearance of components, users may directly compare visual properties, such as size and position, without a careful reference to the scales and attributes of individual components, leading to incorrect insights.
% The repetition visualizations are well suited for listing a group of data with the same data structure.
In summary, we observe two phenomena in visualizations with repetition patterns:
% For repetition visualizations, we  two design suggestions:
\begin{itemize}[$\diamond$]
    \item Keeping the component number between 2 and 8 in a repetition visualization is the most popular. When using it as the main view for analysis, the number is even fewer ($[2,4)$).
    \item In many examples, it tends to omit unnecessary visual elements to reduce visual clutter. For example, using a shared axis and removing duplicated marks to save space and assist comparison (\autoref{fig:juxtaposed_examples}-A5). 
\end{itemize}

\subsubsection{Mirror Patterns}
\setlength\intextsep{0pt}
\setlength{\columnsep}{4pt}
\begin{wrapfigure}{l}{0.05\textwidth}
    \includegraphics[width=0.048\textwidth]{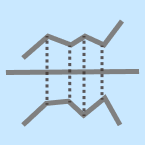}
\end{wrapfigure}
\textbf{Definition}: mirror pattern refers to symmetrically placing two components of the same structure with respect to a coordinate axis. Two components have the same scale on both sides of an axis of symmetry.
Visualizations with mirror patterns are much less popular (2.3\%, 40/\patternamount) in our corpus.
As shown in \autoref{fig:juxtaposed_examples}-B1, for mirror patterns, the most frequent basic type is bar chart (62.5\%, 25).
Other popular types used in mirror visualization include composite visualization (10.0\%, 4), scatterplot (7.5\%, 3), area chart (7.5\%, 3), and heatmap (5.0\%, 2).

Visualizations with mirror patterns mainly have two advantages.
First, taking advantage of people's experience with mirror reflections, mirror visualizations imply identical objects and invite people to compare the two mirrored components~\cite{tufte1985visual}.
% First, with separated layouts, mirror visualizations outperform repetition visualizations on value comparison, especially on shape comparison\highlight{[]}.
%\ww{need to rephrase, since you keep mentioning it is not good for comparison later.}
Second, mirror visualizations are aesthetically pleasing because of their symmetry.
From the samples, we discover that in some cases~\cite{el2018topic, wu2018forvizor}, mirror visualizations are adopted as a part of a symmetrical design, as shown in \autoref{fig:juxtaposed_examples}-B2 \&~\ref{fig:juxtaposed_examples}-B3.
However, mirror visualizations also have two obvious drawbacks.
First, they only support comparing two data series.
Second, because of the symmetry layout, it is more difficult to discover precise differences between components.
Instead, according to the studies by L'Yi et al.~\cite{l2020comparative}, overlaying two data series or using explicit encoding is better than mirror/repetition visualizations in spotting subtle differences.
%The reasons that mirror visualizations are rarely used may be more space required by separated layout but limited ability for comparison.
Therefore, mirror visualizations might not be a good choice when the main design goal is precise comparison, but they can be used as auxiliary components within a symmetric design.
When using mirror visualizations, a number of designs use an explicit encoding to represent the difference~\cite{l2020comparative, srinivasan2018s} or hide the redundant elements.
For example, in ForVizor~\cite{wu2018forvizor}, a soccer analytics system, when visualizing defensive effectiveness, the bars of the offensive team are hidden (\autoref{fig:juxtaposed_examples}-B3).

% \begin{figure}[!htb]
%     \centering
%     \includegraphics[width=\linewidth]{figures/statistics/stacked.pdf}
%     \caption{Distribution of the type combinations in repetition visualizations. Here we only show the rows and columns whose amounts are higher than 1\% of the total amount.}
%     \label{fig:stacked_class}
% \end{figure}

% \begin{figure*}[!htb]
%     \centering
%     \includegraphics[width=\textwidth]{figures/stacked_examples.pdf}
%     \caption{Examples of stack visualizations.}
%     \label{fig:stacked_examples}
% \end{figure*}
\subsubsection{Stack Patterns}
\setlength\intextsep{0pt}
\setlength{\columnsep}{4pt}
\begin{wrapfigure}{l}{0.05\textwidth}
    \includegraphics[width=0.048\textwidth]{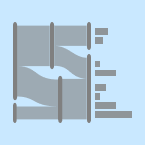}
\end{wrapfigure}
\textbf{Definition}: components of different types or structures are aligned or concatenated by the same data items or a shared margin (e.g., axis) in a stack visualization.
For example, the bar chart + matrix in \autoref{fig:juxtaposed_examples}-C3 and the icicle plot + matrix in \autoref{fig:juxtaposed_examples}-C4 are aligned by items.
The shared margin might not be an axis with the same scale, such as \autoref{fig:juxtaposed_examples}-C5, where the bar charts are stacked together with x-axes representing different levels of data.
Please note that, although a repetition visualization may also have a shared axis, stack visualizations are different in terms of representation and usage.
First, the components in a stack visualization are often different, while components in a repetition visualization are strictly homogeneous.
Second, a repetition visualization is mainly used for listing similar data, while a stack visualization focuses more on presenting different facets of the same data in an interconnected manner.

In our corpus, stack visualizations take up 9.8\% (171/\patternamount).
From the co-occurrence matrix (\autoref{fig:juxtaposed_examples}-C1), we discovered that the most frequent combination is bar chart + matrix.
Among all types, the bar chart is the most frequent (32.5\%, 67).

Going through examples, we identified a common usage that bar charts serve as supporting components to another prominent component with a larger size and a more central position.
The supporting charts are mainly used to show visual summaries of the main component.
For example, a bar chart can summarize the distribution of the data on rows (or columns) of a matrix (\autoref{fig:juxtaposed_examples}-C3).
%Bar charts can also summarize the information of specific axes of bar charts (e.g., LineUp~\cite{gratzl2013lineup}) and scatterplots (e.g., data projections~\cite{meyer2010multeesum, song2016gazedx}).
For other stack visualizations, components often have similar sizes and represent different data aspects (\autoref{fig:juxtaposed_examples}-C4), where there are no leading or supporting roles.
%Please note that the stack visualizations are different from repetition visualizations with shared axes on their visual representations and scenarios.
%repetition visualizations with shared axes are mainly used for visual comparison in a space-efficient manner, but the stack visualizations focus on the relationships and information revealed by different visualization types.

Besides, we observed cases where more than two components can be stacked together to form a large visualization.
In these cases, some components are used as intermediates to connect two or more components.
For example, in \autoref{fig:juxtaposed_examples}-C3, the map connects the heatmaps on the top and to the left.
In addition to the grid layout of components in this example, we also observed a novel linear layout (\autoref{fig:juxtaposed_examples}-C7), where matrix, scatterplot, Sankey diagram, and bar chart are connected together through mutual stacking.
The visual elements of intermediates handle different aspects of data, such as attributes (e.g., coordinates of parallel coordinate plots), data type (e.g., nominal axis and numeric axis of bar charts), data items (e.g., rows and columns of matrices), and data groups (e.g., nodes of Sankey diagrams).
One drawback of linear layout is insufficient space usage when stacking multiple visualizations in different directions (\autoref{fig:juxtaposed_examples}-C8).

Stack visualizations have two major advantages.
First, a stack visualization can present different aspects of the same data at the same time in a compact manner, which conforms the rule of space/time resource optimization in designing multiple-view visualizations~\cite{wang2000guidelines}.
Second, relationships between two stacked components are maintained by a shared margin or shared visual items.
Users can conveniently switch back and forth between components to explore the data because of such visual continuity.
%Users can conveniently switch between components the shared axes and items.
However, the visual continuity is reduced when the number of items or the distance between aligned items increases.
For example, in \autoref{fig:juxtaposed_examples}-C4, although the icicle plot and matrix are adjacent, the icicles with large numbers are distant from the matrix cells, making the alignment and interpretation difficult. 
% However, stack visualizations often suffer from the scalability issue, because the data in different components should be at the same scale\ww{why same scale, and why means scalability issue?}.
We discover the following phenomena based on our observation of stack visualizations:
% and improving the usability and space-efficiency:
\begin{itemize}[$\diamond$]
    \item A visualization could be created by connecting different charts for better visual coherence using intermediate components if the back-end data of the charts are related (e.g., \autoref{fig:juxtaposed_examples}-C8). In this case, stacking multiple visualizations along the same direction or using a grid layout could improve space usage.
    % A novel example is presented in \autoref{fig:juxtaposed_examples}-A5.
    % Here we recommend some visualization types whose visual elements can be easily used to switch between different data attributes . 
    % \item If possible, .
    \item When the number of aligned items is large or the alignments obscure, visual hints are used to indicate the relationships, such as color encoding, highlighting-on-hover, or visual links. For example, the bar charts in \autoref{fig:juxtaposed_examples}-C5 are not strictly aligned, but the color encoding helps to identify the correspondences.
    % Stacking multiple components on different directions may require much more space (\autoref{fig:juxtaposed_examples}-A5).
\end{itemize}

\begin{figure*}[bt]
    \centering
    \includegraphics[width=\textwidth]{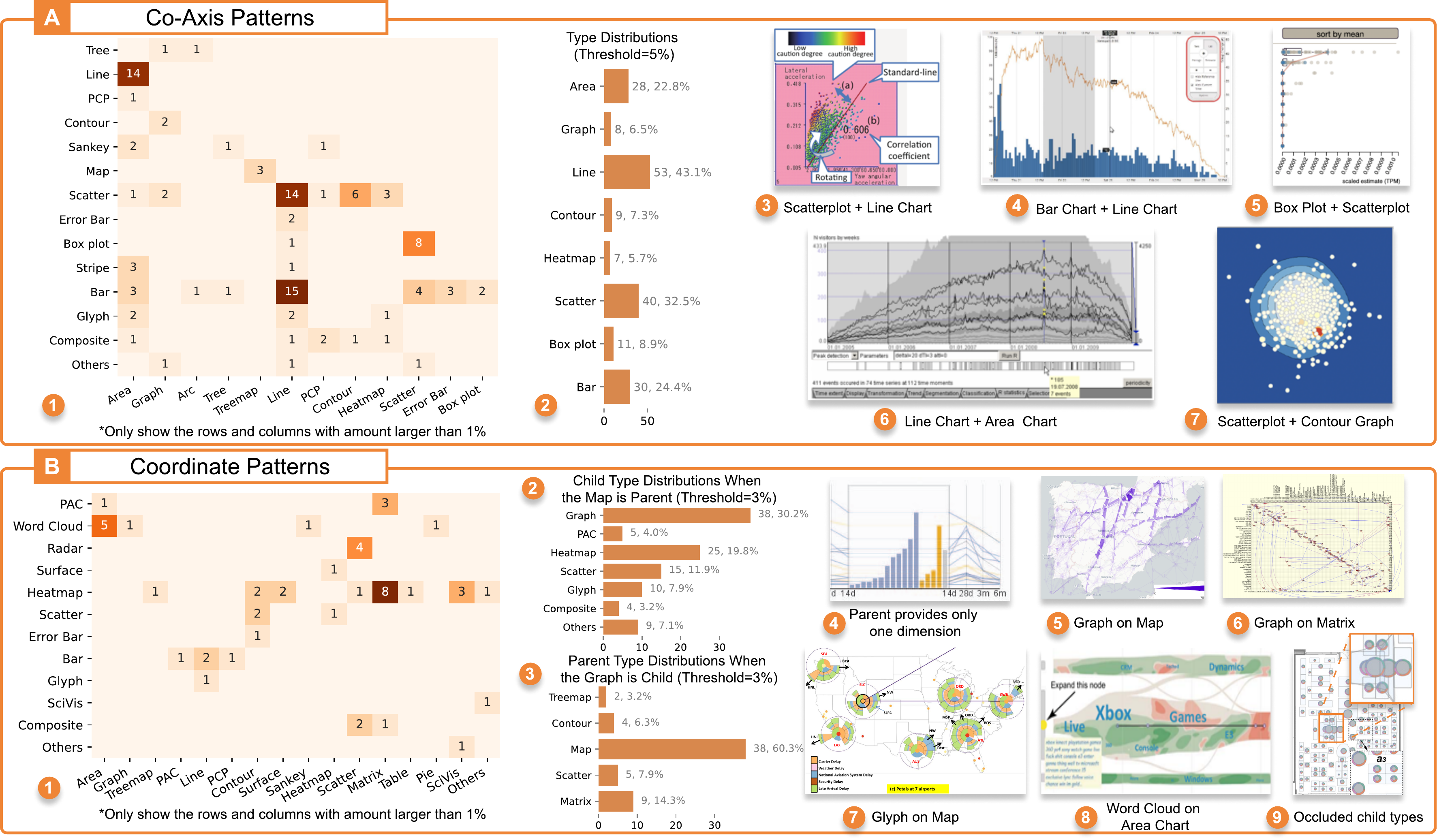}
    \caption{Statistics and examples of co-axis patterns (A: A3~\cite{itoh2015system}, A4~\cite{jo2014livegantt}, A5~\cite{strobelt2015vials}, A6~\cite{andrienko2010discovering}, \& A7~\cite{cao2017voila}) and coordinate patterns (B: B4~\cite{kim2009geneshelf}, B5~\cite{andrienko2018analysis}, B6~\cite{jeong2011state}, B7~\cite{ko2014analyzing}, B8~\cite{wu2014opinionflow}, \& B9~\cite{liu2018tpflow}). 
    }
    % \vspace{-3mm}
    \label{fig:overlay_examples_1}
\end{figure*}
\begin{figure*}[tb]
    \centering
    \includegraphics[width=\textwidth]{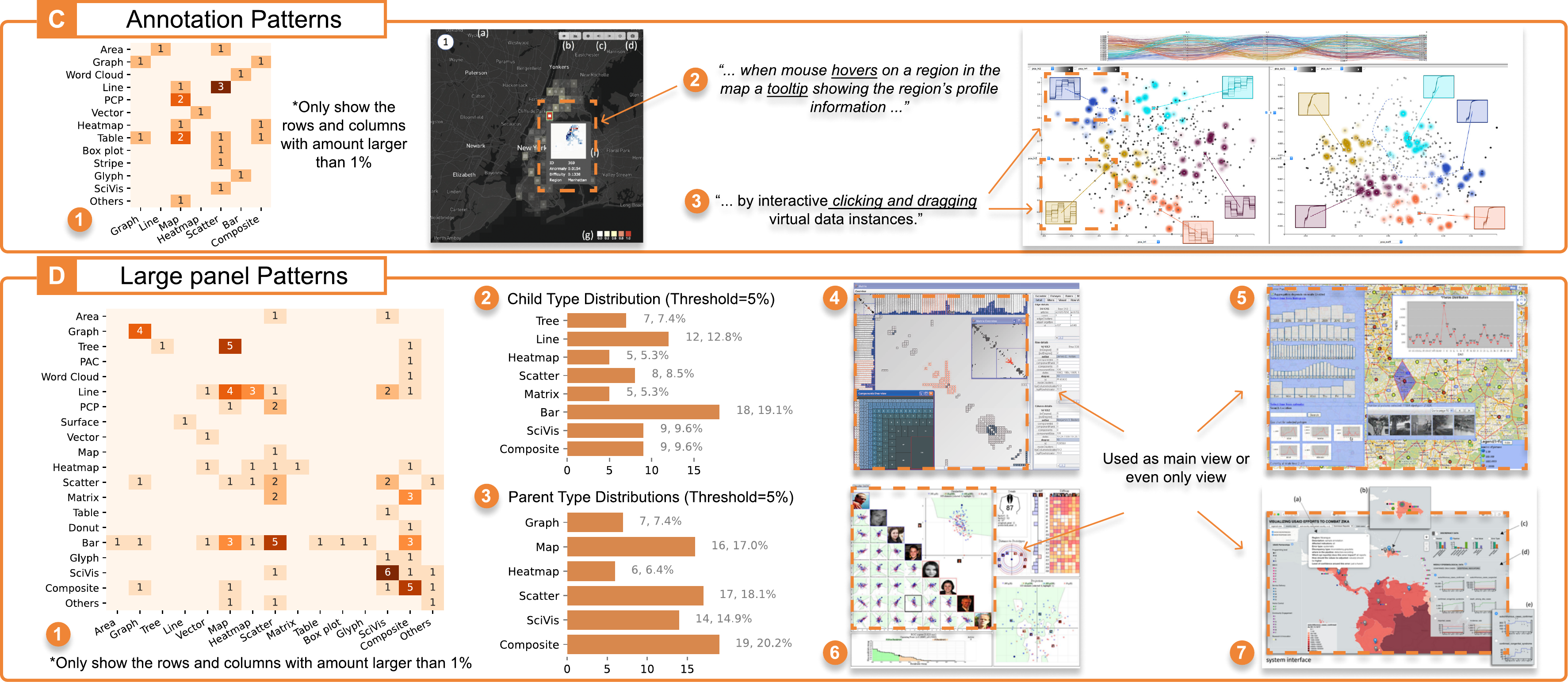}
    \caption{Statistics and examples of annotation patterns (C: C2~\cite{cao2017voila} \& C3~\cite{orban2018drag}) and large panels (D: D4~\cite{henry2006matrixexplorer}, D5~\cite{peca2011kd}, D6~\cite{migut2011diss}, \& D7~\cite{mccurdy2018framework}). 
    % For co-occurrence matrices (C1 \& D1), the rows and columns represent child and parent types, respectively. There are examples of visualizations with annotations (C2~\cite{cao2017voila} \& C3~\cite{orban2018drag}) and large-panel visualizations (D4~\cite{henry2006matrixexplorer}, D5~\cite{peca2011kd}, D6~\cite{migut2011diss}, \& D7~\cite{mccurdy2018framework}).
    }
    % \vspace{-3mm}
    \label{fig:overlay_examples_2}
\end{figure*}

\subsection{Overlay}
Visual components are overlaid over other components in an overlay visualizations.
Overlay visualizations take up 23.7\% (415/\patternamount) of the collected examples.

Overlay visualizations have two advantages.
First, a visualization with overlay patterns often has a more compact layout compared with juxtaposition.
Second, overlay patterns can directly represent the correspondences between different components, thus enhancing the visual effect.
However, a common disadvantage of overlay patterns is occlusion when compared with juxtaposition visualizations and nesting visualizations.
Therefore, when designing an overlay visualization, it would be better to use clutter reduction techniques (e.g., edge bundling) to improve the visual appearance.
Overlay visualizations can be organized in four categories: \textit{co-axis}, \textit{coordinate}, \textit{annotation}, and \textit{large panel}.
A co-axis visualization contains multiple visualizations that share the same coordinate system, while the other three categories all refer to cases that smaller visualizations (\textbf{child components}) are overlaid on the top of larger visualizations (\textbf{parent components}).

\subsubsection{Co-Axis Patterns}
\setlength\intextsep{0pt}
\setlength{\columnsep}{4pt}
\begin{wrapfigure}{l}{0.05\textwidth}
    \includegraphics[width=0.048\textwidth]{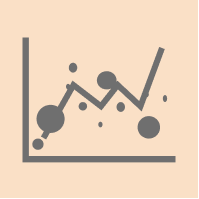}
\end{wrapfigure}
\textbf{Definition}: component visualizations share the same coordinate system in a co-axis visualization.
Co-axis visualizations take up 6.9\% (121/\patternamount) in our corpus.
\autoref{fig:overlay_examples_1}-A2 shows that the most frequent type include line chart (43.1\%, 53), scatterplot (32.5\%, 40), area chart (22.8\%, 28), and bar chart (24.4\%, 30).
From the co-occurrence matrix (\autoref{fig:overlay_examples_1}-A1), we observe that the top five combinations are bar chart + line chart (15), scatterplot + line chart (14), line chart + area chart (13), box plot + scatterplot (8), and scatterplot + contour graph (6).
Going through the samples, we discover several specific tasks for these common combinations.
First, scatterplots, which show detailed data items, are often combined with other summary visualizations of trends (line charts), distributions (box plots and contour graphs), etc.
% The type scatterplots can show detailed data items while composing with other types to show the overall trend (line charts) or distribution (box plots for 1D distribution and contour graphs for 2D distribution).
The combination of area chart and line chart exhibits various usages, such as showing uncertainty or differences of the lines with area chart~\cite{hao2009poster}, representing aggregated lines with areas~\cite{XuMR017} to reduce visual clutter, or using area charts as a special case of line charts~\cite{wang2018dqnviz}.
The bar chart + line chart is adopted to visualize independent data series in most cases~\cite{jo2014livegantt, li2015fpsseer,li2020cnnpruner, zhong2020silkviser}. In rare cases, line charts serve to show density plots for bar charts~\cite{chen2015interactive}.
In particular, we discovered 4 out of 15 cases in which bar charts and line charts likely share the same coordinate systems, but they actually have different y-coordinates, which would be easily overlooked (e.g., \autoref{fig:overlay_examples_1}-A4).
% ~\cite{li2020cnnpruner, zhong2020silkviser, jo2014livegantt}.
% Line charts are commonly used to show the regression or trend of the scatterplots (called trend line in some cases~\cite{borkin2013makes}); and scatterplots can show the detailed data items, and box plots (1-D) and contour graphs (2-D) can show the overall distribution.

The advantage of co-axis patterns is that placing multiple components in the same coordinate system facilitates direct comparison and pattern recognition.
We obtain two observations considering the occlusions between components.
\begin{itemize}[$\diamond$]
    % \item Use accompanied views under specific tasks that the users have an intuition for what different visualizations are used for, e.g., regressing (scatterplot and line chart), finding anomaly (box plot and scatterplot), visualizing uncertainty (bar chart and error bar).
    \item Overlaid components might use transparency to reduce occlusion or put summary/important components on the top. For example, placing box plots on top of a scatterplot for anomaly detection tasks.
    \item A number of designs adopt multiple coordinate systems in a co-axis visualization, which might introduce potential biases~\cite{isenberg2011study}. For example, the bar chart and line chart in \autoref{fig:overlay_examples_1}-A4 use different y-coordinates, so that users may misinterpret.
\end{itemize}

%For these design patterns, rows of the co-occurrence matrix represent the child types and columns represent the parent types.

\subsubsection{Coordinate Patterns}
\setlength\intextsep{0pt}
\setlength{\columnsep}{4pt}
\begin{wrapfigure}{l}{0.05\textwidth}
    \includegraphics[width=0.048\textwidth]{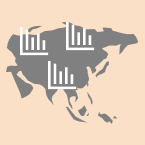}
\end{wrapfigure}
\textbf{Definition}: in a visualization design with coordinate design patterns, parent components provide coordinates (e.g., Cartesian coordinate system, geographic coordinate system, and other reference systems such as grids of the matrix) for child components (or their visual elements). The reference systems are regarded to be part of the parent components.
In other words, the positions of child components encode back-end spatial data referring to the parent component.
For example, in \autoref{fig:old_taxonomy}B, the map provides spatial context for the graph nodes.
Similarly, in \autoref{fig:old_taxonomy}C, the positions of the graph nodes are determined by the treemap grids.
Please note that, in some cases, parent components may only provide x- or y-positions for child components.
For example, in \autoref{fig:overlay_examples_1}-B4, the parallel coordinates plot (PCP) provides one of its axes as a reference to the bar chart.
Another example is embedding scatterplots into PCP (\autoref{fig:old_taxonomy}D), where one axis is rotated 90 degrees, creating a 2D coordinate system with another axis to host the scatterplots.

In total, we obtained 196 (11.2\%/\patternamount) samples with coordinate patterns.
% \autoref{fig:overlay_examples_1}-B1 shows the co-occurrence matrix of visualization types in these samples, where the rows represent the child types and columns represent parent types.
% For the rest design patterns, we used similar matrices to visualize and analyze the relation between components.
Among the samples, 63 (31.3\%) have graphs as child components and 126 (62.7\%) contain maps as parents, making the type co-occurrence matrix (\autoref{fig:overlay_examples_1}-B1) highly sparse.
Therefore, we separate them from the type co-occurrence matrix and visualize their type distributions independently (\autoref{fig:overlay_examples_1}-B2 \& \ref{fig:overlay_examples_1}-B3).
Among all combinations, overlaying graphs on maps is the most frequent.

Graphs are mostly used as child components (31.3\%, 63).
It is likely because, in many graph visualization tasks, analysts mainly focus on topological patterns, such as node degree and node connectivity~\cite{lee2006task}. 
Therefore, node positions are relatively flexible depending on the needs of different tasks, and encoding spatial information with node positions becomes a popular design pattern.
%For example, to make graphs aesthetically pleasing~\cite{bennett2007aesthetics}, e.g., with circular layouts, symmetric layouts, or less edge clutter, the node positions can be adjusted.
%In addition, we also observe that maps mostly used as parent components.
In addition, there are many cases with word clouds referring to area charts (\autoref{fig:overlay_examples_1}-B8) or proportional area charts referring to  matrices.
In these cases, child components regularly do not present spatial information.
%Therefore, we conclude a design strategy towards coordinated visualizations is by providing spatial context the visualizations whose visual elements do not encode spatial information.

Compared with co-axis patterns where the components have independent but identical coordinate systems, the layout of child components is determined by their parent components in coordinate patterns.
Therefore, they are effective in helping users interpret child components in the context of a parent component.
% However, coordinated visualizations require sufficient space for child components to provide detailed information to the points of interest of parent components.
We discover two phenomena with coordinate patterns.
\begin{itemize}[$\diamond$]
    \item Various designs choose to combine a parent component that provides spatial context and child components whose visual elements do not encode spatial information, such as word clouds and proportional area charts. For example, in \autoref{fig:overlay_examples_1}-B8, the words in the word clouds are distributed on an area chart to visualize the topic frequency.
    \item In addition to the inevitable occlusions between child and parent components, the overlapping between the child components may exacerbate the overall occlusions. For example, in \autoref{fig:overlay_examples_1}-B9, the glyphs are used to enhance the map visualization, but users might fail to retrieve the information encoded by glyphs because of occlusion.
    % \item Carefully select visualizations with high visual occlusion as child components, because the coordination may worsen the visual occlusion. \highlight{example?}
\end{itemize}

% In addition to coordinated visualizations, there are cases that parent components do not provide coordinates for child components.
% These cases are called annotation and large panel according to Bach et al~\cite{bach2018design}.

\subsubsection{Annotation Patterns}
\setlength\intextsep{0pt}
\setlength{\columnsep}{4pt}
\begin{wrapfigure}{l}{0.05\textwidth}
    \includegraphics[width=0.048\textwidth]{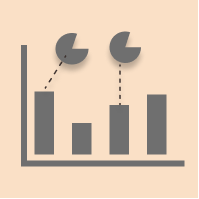}
\end{wrapfigure}
\textbf{Definition}: child components of small size are overlaid on parent components and connected to elements of parent components in annotation visualizations, but the positions of child components do not encode spatial information. Moreover, the child components provide a ``cut-out'' lens for the visual elements connected~\cite{bach2018design}.
Compared with co-axis and coordinate, in which components are related to each other because of sharing or providing coordinate systems, annotation visualizations are more flexible as child components have more freedom in placement and can use visual links to explicitly connect to the parent component.
There are only 22 (1.3\%) annotation visualizations in our corpus.
% \autoref{fig:overlay_examples_2}-C1 shows the type co-occurrence.

% From type co-occurrence matrix (\autoref{fig:overlay_examples_2}-C1), we observed that the most common combination is line chart + scatterplot.
%However, the co-occurrence number may not imply significance because of small sample size.
We reviewed the captions and text descriptions in the corresponding papers to understand the scenarios of annotation visualizations.
17 out of 22 have mentioned that the child components are displayed on demand (via interactions with the parent components).
%Among all samples, we discovered 12 have explicitly mention that after interacting on parent components, and child components will pop out then.
Therefore, the most common usage of annotation patterns is showing additional information with tooltips~\cite{eick2007thin, guo2017eventthread, cao2017voila}.
For example, in \autoref{fig:overlay_examples_2}-C2, when hovering on a grid on the map, a graphical annotation about the profile of that grid will present.

The advantage of annotations is the flexibility in positioning child components.
% However, when the number of the child components gets large, the organization of child components will be challenging.
We discover the following two phenomena of annotation visualizations.
\begin{itemize}[$\diamond$]
    \item In most cases, only details of focused data items are visualized following the rule of details on demand~\cite{shneiderman2003eyes}.
    \item The layout of child components can be optimized (e.g., saliency-based method~\cite{steinberger2011context}) to utilize empty space or reduce line crossings.
\end{itemize}
%Besides, there are cases that child components are not linked to parent components. For these cases, we identified large-panel visualizations.

\subsubsection{Large Panel Patterns}
\setlength\intextsep{0pt}
\setlength{\columnsep}{4pt}
\begin{wrapfigure}{l}{0.05\textwidth}
    \includegraphics[width=0.048\textwidth]{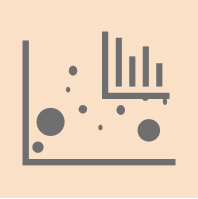}
\end{wrapfigure}
\textbf{Definition}: child components of small size overlay directly on parent components without visual links in a large panel visualization, and the positions of child components do not encode spatial information.
Unlike annotations, large panels do not connect the child and parent components using links or anchors, and the child components show details of the parent components in an overview + detail manner~\cite{bach2018design}.

% Compared to annotation visualizations, the size of the child components of the large-panel visualizations may be larger. 
In total, We obtained 76 (4.3\%) samples of large-panel visualizations.
% The co-occurrence matrix (\autoref{fig:overlay_examples_2}-D1) 
\autoref{fig:overlay_examples_2}-D2 \&~\ref{fig:overlay_examples_2}-D3 show the distributions of visualization types used as child and parent components, respectively.
By exploring the samples, we found that large-panel visualizations are mostly used as the main views in visual analytics systems, even as the only view in some systems (\autoref{fig:overlay_examples_2}D).
Child components generally serve as auxiliary views for the whole parent components, not specific elements of the parent components.
This feature makes large-panel visualizations different from annotation visualizations, where child components usually present the details of elements in parent components.

Compared with annotation patterns, large-panel visualizations offer more flexibility for placing child components, since they do not require anchoring points in the parent component.
% On the other hand, large-panel visualizations generally hold fewer child components than annotation visualizations.
% In most cases, we only observed one or two child components.
% The advantage of large-panel visualization is flexible positions of child components, which is similar to annotation visualizations.
For large-panel visualizations, they generally place child components at positions where elements are less important (such as corners) to mitigate visual occlusion.
\begin{figure*}[!htb]
    \centering
    \includegraphics[width=\textwidth]{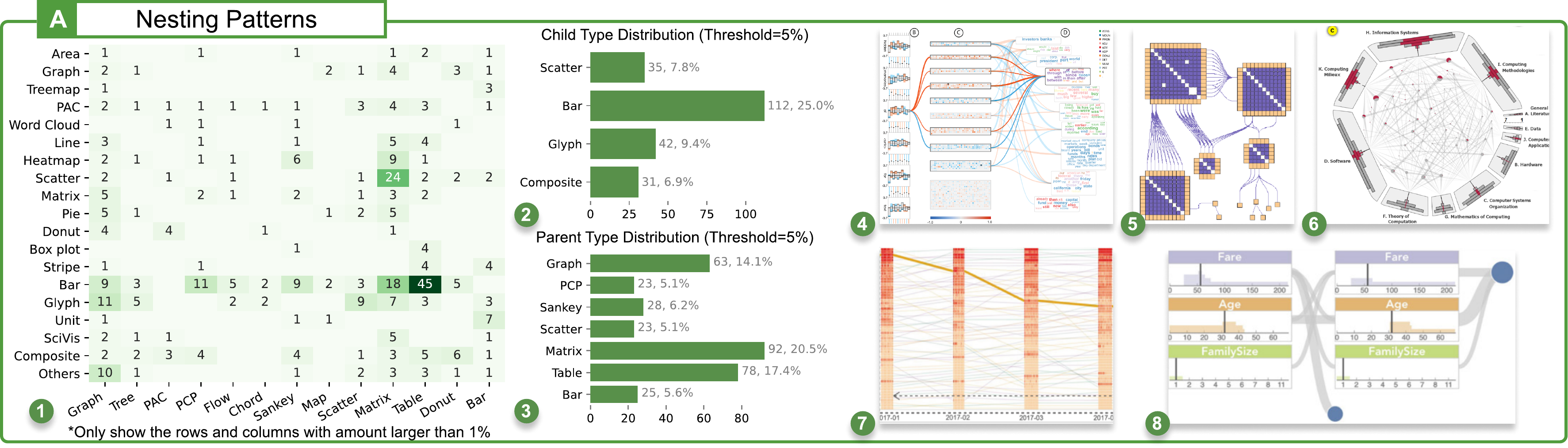}
    \caption{Statistics and examples of nesting visualizations (A4~\cite{ming2017understanding}, A5~\cite{y2008improving}, A6~\cite{alsallakh2013radial}, A7~\cite{yue2018bitextract}, A8~\cite{zhao2018iforest}).}
    % \vspace{-3mm}
    \label{fig:nested_examples}
\end{figure*}

\subsection{Nesting}
\setlength\intextsep{0pt}
\setlength{\columnsep}{4pt}
\begin{wrapfigure}{l}{0.05\textwidth}
    \includegraphics[width=0.048\textwidth]{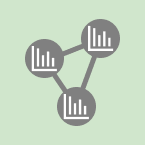}
\end{wrapfigure}
\textbf{Definition}: in nesting visualizations, some components (denoted by \textbf{child components}) are embedded into the visual elements or internal area of other components (denoted by \textbf{parent components}).
We collected 392 (22.4\%) samples of nesting visualizations in total.
The co-occurrence matrix (\autoref{fig:nested_examples}-A1) shows that nesting visualizations have more diverse type combinations than other composition patterns.
The most frequent combinations include scatterplots + matrix (a.k.a scatterplot matrix) and bar chart + table.
Bar charts and scatterplots are the most common child type.
For parent components, graphs, matrices, tables, Sankey diagrams, bar charts, parallel coordinate plots (PCP), and scatterplots are significantly more popular than other types.
We also observed different patterns of visualizing child components in nesting visualizations.
Small elements of parent components, such as nodes of graphs (\autoref{fig:nested_examples}-A5), nodes and flows of Sankey diagrams (\autoref{fig:nested_examples}-A4), sectors of donut chart, and cells of matrix and table are obvious visual spaces to embed child components.
However, some other parent visual elements, such as axes of PCP, need to distort to create a canvas to host child components (\autoref{fig:nested_examples}-A7).
In addition, there are nesting visualizations where parent components have circular shapes and internal area, such as donut (\autoref{fig:example} \&~\ref{fig:nested_examples}-A6). 
The internal area can provide relatively sufficient space other than visual elements.
These cases are not common (6.4\%, 25) among nesting visualizations.
% These visualizations usually have only one child component, which might connect directly (10/25), align along the radius (6/25), or without explicit connection (9/25) to the parent components. 

Nesting visualizations have two advantages.
First, they have no occlusions between parent and child components and imply hierarchical information, compared to overlay visualizations.
Therefore, nesting visualizations can visualize the overview of parent components (e.g., the overall layout of graphs) while maintaining details of the child items (e.g., graph nodes and matrix cells)~\cite{elmqvist2009hierarchical}.
Second, nesting visualizations are more compact than overlay visualizations and juxtaposition visualizations.
However, one major limitation of nesting visualizations is the limited space of visual elements to host child components. 
%a concern to use nesting visualizations is that limited visual space for child components may make the details unseeable.
From the observations, we discover two phenomena.
\begin{itemize}[$\diamond$]
    \item A number of designs choose to use relatively common/simple visualizations in child components, such as bar charts (\autoref{fig:nested_examples}-A8) and heatmaps (\autoref{fig:nested_examples}-A4 \&~\ref{fig:nested_examples}-A7). We infer that this is because visualizations with complex configurations are hard to identify due to the limited space of child components.
    % Only show summary information in child components. Too much detailed information will overwhelm users. For example, bar charts (\autoref{fig:nested_examples}-A8) and heatmaps (\autoref{fig:nested_examples}-A4 \& \ref{fig:nested_examples}-A7) are popular choices for child components.
    \item A number of designs apply geometric transformations to the elements of parent components to make room for child components. For example, Sun et al.~\cite{sun2016embedding} proposed a route-zooming technique to distort the map for hosting visualizations for spatio-temporal information.
\end{itemize}
\section{Usage Scenario}
\label{sec:secnario}
The taxonomy and corpus can be used in different aspects. 
% \subsection{Design Inspiration \& Understanding}
\subsection{Exploring \& Understanding Visual Designs}

\begin{figure*}[!htb]
    \centering
    \includegraphics[width=\linewidth]{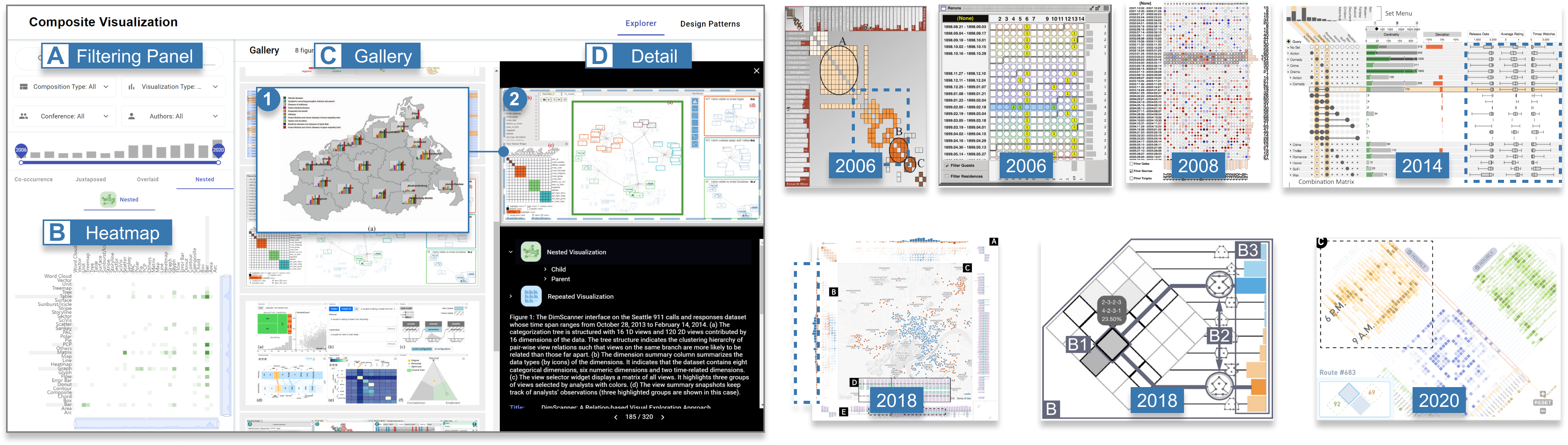}
    \caption{Left: composite visualization explorer with a filtering panel (A), a heatmap view (B), a gallery (C), and a detail view (D). Right: stack visualizations with matrices and bar charts across different years.}
    % \vspace{-3mm}
    \label{fig:explorer}
\end{figure*}

% As demonstrated by Munzer~\cite{munzner2009nested} and Satyanarayan et al.~\cite{satyanarayan2019reflection}, visualization design is conducted iteratively by considering data, tasks, visual representation, and composition.
% For designers, the taxonomy and corpus can help them design by providing information about visual representations and compositions.
The corpus can help researchers and designers in explore and understand composite visualization designs.
To facilitate design exploration, we developed an explorer for composite visualizations.
The explorer consists of four components: a \textbf{filtering panel} (\autoref{fig:explorer}-A), a \textbf{heatmap view} (\autoref{fig:explorer}-B), a \textbf{gallery view} (\autoref{fig:explorer}-C), and a \textbf{detail view} (\autoref{fig:explorer}-D).
The filtering panel supports filtering designs by keywords, year, venue, composition pattern, and visualization type.
After filtering, an overview of type combinations under different composition patterns is displayed in the heatmap view.
Each grid in the heatmap encodes the number of a type combination.
The heatmaps support filtering designs of specific type combinations by clicking on the corresponding cells.
The filtered designs are displayed in the gallery view.
When clicking on a design in the gallery view, a window pops up to show details about that design, including our annotations and metadata, such as the title.
The interface facilitates searching for a visualization type with different composition patterns.
For example, \autoref{fig:explorer}-D1 shows bar charts being displayed on a map visualization with the coordinate pattern, and \autoref{fig:explorer}-D2 nests bar visualizations into the graph nodes.

Moreover, using the interface to display visualizations of specific type combinations and composition patterns side by side enables understanding the evolution of the design. For example, we can witness an evolution of design complexity for the visualizations with stacking bar charts and matrices from \autoref{fig:explorer} (Right).
In earlier years (2006 and 2008), designs differentiate from each other in terms of shape and color encodings of the matrix cells.
Then more different visual elements are concatenated with bar charts, such as boxplots (2014).
Composite visualizations of bar charts and matrices can further serve as support components for map visualization (2018).
Besides, there are different directions for matrices when stacking with bar charts.
In a more recent design (2020), bar charts are stacked with multiple matrices in a crossing/exploding layout.

% Before designing a new visualization, a designer might have a rough idea about the choice of visualizations. For example, specific columns of a dataset are suitable to use a bar chart for visualization. Then, s/he has two choices. The first one is starting from the taxonomy, which could serve as a reference that how a bar chart is used in different patterns, such as being a parent component in a coordinated visualization. However, from the statistical analysis, s/he discovers that the bar chart is not frequently used as a parent component but as a child component, which might influence her/his design choices. Furthermore, the advantages, disadvantages, and observed usages of the design patterns can help compare different designs. Second, s/he could explore the corpus by searching for composite visualizations that contain a bar chart, and inspect how practical designs organize bars with other visual components to represent more data attributes.

\subsection{Training Data for AI4VIS}
% For researchers, our work can inspire potential studies in different research directions. First, for empirical studies, our descriptions of the observations on different design patterns leave much room for proposing hypotheses and designing experiments for validation. For example, we discovered that in stacked visualizations, specific types could serve as supporting components for the main view. Existing theories of multiple-view visualizations (e.g., levels of abstraction~\cite{wang2000guidelines}) could partially explain their relations. Nevertheless, it is required for further exploration that what kind of tasks the supporting components can achieve and how effective the different types are. 

Our corpus can be used as training data for artificial intelligence models for visualizations (AI4VIS)~\cite{wu2021survey}. A possible task is decomposition, which comprises two sub-tasks, i.e., recognizing positions and types of basic visualizations and inferring the composition patterns of the visualizations.
The bounding boxes and labels can be used for visualization detection~\cite{chen2021vis30k, deng2021visimages}.
For the visualizations assigned with multiple labels (e.g., heatmap and map), the data is represented with multiple bounding boxes with the same x, y, width, and height but different labels. 
VisImages~\cite{deng2021visimages} has demonstrated a case for this situation.
With bounding boxes and visualization types, we can further infer the composition patterns, which describe the relationships between the basic visualizations. Recent studies propose practical methods to learn the hierarchical structures of visual elements with graph neural networks~\cite{li2022structure} or transformer-based model~\cite{shi2022reverse}.
Furthermore, our annotation comprises the co-occurrence between basic visualizations, which could be used for visualization recommendation~\cite{vartak2017towards} based on knowledge graph (e.g., basic visualizations as the entities and design patterns as the relations).
\section{Future Research Opportunities}
In this section, we discuss future research opportunities for implementing composite visualizations and exploiting empirical evidence for task-driven efficiency.
% \textbf{From perspective of coordination.} The use of coordinates is different under the stacked, accompanied, and coordinated visualizations.
% % In stacked visualizations, 
% 1) data annotation.
% 2) taxonomy.
% 3) findings.
% \noindent{\textbf{Metaphors of Composition Patterns.}}
% Juxtaposed, overlaid, and nesting visualizations differ from each other on overlapping relations between the components.
% Visualizing more than one components in an individual visual space improves information density 
% In general, a more complex design posts higher requirements on skills of designing and implementation.
% Similarly, a more compact design has higher information density, with higher potential to create visual clutter.
% \textit{Visual Effects vs. Visual Clutter.} Visual clutter refers to confusion and disorder because of full of visual objects.
% Issues to resolve:
% \begin{itemize}
%     \item stacked definitions, drawbacks, and suggestions
%     \item accompanied suggestions
%     \item coordinated suggestions: strategies for creating novel coordinated visualizations
%     \item nested cases
%     \item names of different types: coordinated? located? positioned? navigated?
%     \item number of nested
%     \item design suggestions: 
% \end{itemize}

\subsection{Implementing Composite Visualizations}
Our taxonomy can be used to measure the expressiveness of existing visualization grammar in visualization rendering and facilitate the development of more ease-of-use visualization generation grammar.
Grammars that support operating on low-level visual elements, such as D3.js~\cite{bostock2011d3} and Vega~\cite{2016-reactive-vega-architecture}, can implement various composition patterns through programming, but they require high programming capability of the users.

In recent years, declarative programming languages are developing rapidly and gradually supporting the generation of visualizations through the intuitive specification of the visual encodings.
For example, Vega-Lite~\cite{satyanarayan2016vega} supports view composition with operators of ``facet/repeat'' (repetition), ``layer'' (part of coordinate patterns and co-axis patterns), and ``vconcat/hconcat'' (part of stack patterns).
However, it cannot generate nesting visualizations when there is a need to represent additional information in the visual elements, such as glyph visualizations~\cite{borgo2013glyph}.
The generation of nesting visualizations also requires the support of processing network/hierarchical data.
ECharts~\cite{li2018echarts} supports graph and tree visualizations.
GoTree~\cite{li2020gotree} facilitates the rendering of hierarchical data by specifying coordinate systems, visual elements, layout, etc. 
However, these grammars lack original support for composition.
ATOM~\cite{park2017atom} supports generating visualizations with nested data units, but the building blocks of generated visualizations are not visualizations.
Nevertheless, its graphical operations, such as bin, duplicate, and filter, can be extended to support generating nesting visualizations, for example, aggregating the transformed data units and rendering them with basic visualization types.
In all, analyzing existing visualization grammars with the taxonomy, we understand that existing grammars can be further extended to support the convenient generation of more composition patterns, especially nesting patterns, which account for 24.2\% in our corpus.

\subsection{Empirical Evidence for Task-Driven Efficiency}
Composition patterns have been applied in many visual analytics systems, which are designed to achieve various analysis tasks~\cite{brehmer2013multi}.
However, the efficiency of visualization composition has long been discussed and the design complexity remains a problem in visualization research~\cite{chen2019ontological, robertson2009complexity, wu2022defence}.
However, directly comparing the design of different visual analytics systems might be impractical due to their complexity.
Our taxonomy provides a breakdown of composition for evaluating the efficiency of complex designs under different tasks~\cite{amar2005low, brehmer2013multi}, such as comparing values and discovering anomalies.

Some studies have investigated composite visualizations for the task of comparison.
% Existing studies have reviewed and studied how different composition patterns affect the efficiency of comparing values.
For example, Isenberg et al.~\cite{isenberg2011study} studied how dual-axis charts, a special co-axis pattern, perform in the tasks of comparing lengths and distances.
L'Yi et al.~\cite{l2020comparative} have thoroughly explored the effectiveness of repeated, mirrored, and co-axis layouts for comparison.

In addition, we observe the use of composite visualizations for other tasks, such as co-axis patterns for discovering correlations/anomalies and coordinate patterns for providing spatial information.
Saket et al.~\cite{saket2018task} have studied how basic visualization types perform in low-level tasks (e.g., finding anomalies, finding clusters, and correlation)~\cite{amar2005low}.
However, few studies have investigated the efficiency of composition patterns for these tasks.
Future studies can conduct controlled experiments in exploring the efficiency of several representative visualizations with compositions.

Nesting visualizations are well suited for representing the network and hierarchical data~\cite{elmqvist2009hierarchical}. Still, these data are usually analyzed under tasks different from tabular data, such as perceiving the topology and the attributes on nodes or links~\cite{lee2006task}.
Existing studies have investigated the perception efficiency of graphs in different conditions, such as static or dynamic manners~\cite{farrugia2011effective} and multiple sampling models~\cite{zhao2020preserving}.
However, few user studies have been conducted to evaluate the efficiency of nesting layouts.
Future studies can design different perception tasks for the child components and parent components in the visualizations and conduct controlled experiments accordingly.
%In these cases, visual components should be coherently organized, thus fulfilling the needs of visual analytics and minimize visual clutter.
% Therefore, designing overlaid and nesting visualizations requires understanding of the tasks and data.
% We hope our corpus and insights can help advance the community from the perspectives of assisting designs.

\section{Discussion}
In this section, we discuss trade-offs of different composition patterns and limitations of composition representations.

\subsection{Balancing Expressiveness and Effectiveness}
Designing a visualization should handle the trade-offs between representing more information in a limited visual space and ensuring users are not overwhelmed by too many visual components~\cite{munzner2014visualization}.
Specifically, juxtaposition can provide flexible layouts for charts with small occlusions, which might be friendly for design novices.
Among juxtapositions, stack patterns can express different aspects of data with a more coherent arrangement. 
Furthermore, overlay and nesting provide more compact layouts than juxtaposition visualizations to handle more visual components at once.
These compositions can also make convenient to perceive spatial, networking, or hierarchical relation between child and parent components.
However, these patterns increase the visual occlusion and limit the size of child components.

\subsection{Coverage of Composition Representations}
We encountered that some designs are not decomposable with our composition representations.
%In these cases, we failed in recognizing the basic types of the components and labeled them as ``others'' type.
For example, Bubble Sets~\cite{collins2009bubble} are highly customized with primitive shapes (such as rectangles, lines, and circles), instead of combining multiple types together.
Moreover, the composition patterns might not fully reflect the design novelty.
For example, in addition to coordinate and nesting, Whisper~\cite{cao2012whisper} use a metaphor of sunflower to represent the retweeting activities.
% coordinates a graph on a map visualization.
% In each graph node, small visual elements representing twitters are nested. 
% For the graph layout, the authors use a metaphor of sunflower to represent the retweeting activities. Another example is scattering points in parallel coordinates (Fig.~\ref{fig:old_taxonomy}D~\cite{yuan2009scattering}), in which dimension reduction and layout optimizing techiniques are used to integrate scatterplots and parallel coordinates.
To understand and analyze these novel designs, lower-level decomposition, which concerns transformations and visual encodings, is required, such as the component layout (e.g., circular and branched).

\subsection{Limitations}
Our study has two limitations.
First, data and interactions of designs, which are closely related to analytical tasks and design requirements, are not considered in our corpus.
The acquisition of this information requires extensive efforts on paper reading and is even inaccessible sometimes.
In this work, we mainly study design patterns regarding observable information in figures, such as visualization types and spatial relationships, because of the large size of the corpus.
A potential solution to retrieve information about data and interactions might be using natural language processing techniques to extract and analyze related text descriptions.
% Third, the coverage of our corpus may not comprehensive enough, leaving some compositions of visualizations unexplored.
% The limited coverage may results in the blank cells in composition matrices.
% Therefore, we will extend our corpus by incorporating designs from more venues (e.g., TVCG, PacificVis, EuroVis, and CHI).
Second, the corpus construction mainly relies on manual annotation, which is limited in scalability for a larger quantity of data.
Recently, studies~\cite{chen2021vis30k, deng2021visimages} have adopted object detection models to process visualization images, which might be a promising method for data collection.
In this work, we lacked a well-curated dataset for model training at the beginning, but we could use the collected corpus to explore the potential of automatically recognizing visualizations and composition patterns.

\section{conclusion and future works}
In this work, we opted to answer the question of what and how visualizations can be composed together to form novel designs.
To achieve this, we conducted a demographic analysis on composite visualizations, based on a corpus of visual designs from IEEE VIS.
With the corpus, we proposed a taxonomy of eight design patterns.
For each design pattern, we analyzed the distributions and correlations between different visualization types, and obtained insights on usage scenarios, advantages, disadvantages, and design suggestions.
% With the analysis, we delivered a holistic view of state-of-the-art composite visualizations, which might inspire future research topics and designs.
We released the corpus and an explorer to advance the studies in designing composite visualizations: \url{https://composite-visualizations.github.io/}.

For future research, one promising direction is a library that can flexibly integrate different visualization types with different composition patterns.
Existing libraries (e.g., Vega-lite, ECharts) can provide support for layering or faceting.
Nevertheless, researchers create composite visualizations mainly by writing codes with programming languages (e.g., Javascript), since composite visualizations are with complex structures and are commonly integrated with visual analytics systems.
With the taxonomy and corpus obtained in this work, we might extend the features of existing libraries for better creation of composite visualizations.
% Existing libraries (e.g., Vega-lite, ECharts) mainly work for the rendering and interactions of basic charts (e.g., bar charts), lacking supports for complex compositions.
% Other authoring tools support creating complex visualizations, but aim to create infographics (such as data factsheets and posters) instead of visual analytic systems.
% Therefore, one promising future direction is a library that can flexibly integrate different visualization types with different composition patterns.
Another research problem is the effectiveness of composition patterns on different tasks.
With the overview provided before, we understand the scenarios and tasks for different design patterns, but in-depth inspections, such as controlled studies, are beyond the scope of this paper.
Recently, L'Yi et al.~\cite{l2020comparative} have explored the effectiveness of different chart compositions (e.g., superimposed and explicit encoding) on visual comparison. 
We hope our taxonomy and findings could shed light on exploring further empirical studies with more diverse composition patterns, visualizations, and tasks.
Furthermore, we could study the interaction techniques in composite visualizations. Specifically, we could use natural language processing approaches with existing interaction taxonomies~\cite{gotz2009characterizing} to retrieve a corpus of interactions in the designs. Combining with the corpus in this work, we might discover the relations between interactions, visualization types, and composition patterns in the works of visualization community.

\ifCLASSOPTIONcompsoc
  \section*{Acknowledgments}
\else
  \section*{Acknowledgment}
\fi

The work was supported by NSFC (62072400) and the Collaborative Innovation Center of Artificial Intelligence by MOE and Zhejiang Provincial Government (ZJU). This work was also partially funded by the Zhejiang Lab (2021KE0AC02).

\ifCLASSOPTIONcaptionsoff
  \newpage
\fi



\bibliographystyle{IEEEtran}
\bibliography{main}


% \end{thebibliography}


\begin{IEEEbiographynophoto}{Dazhen Deng}
  is currently a Ph.D. student in the State Key Lab of CAD\&CG, Zhejiang University.
  He received his B.E. degree from Zhejiang University in 2018.
  His research interests mainly lie in machine learning for visualization and sports visualizations. For more information, please visit https://dengdazhen.github.io.
  \end{IEEEbiographynophoto}
  
  \begin{IEEEbiographynophoto}{Weiwei Cui}
    is a Principal Researcher at Microsoft Research Asia, China. He received his PhD in Computer Science and Engineering from the Hong Kong University of Science and Technology and his BS in Computer Science and Technology from Tsinghua University, China. His primary research interests lie in visualization, with focuses on text, graph, and social media. For more information, please visit http://research.microsoft.com/en-us/um/people/weiweicu/.
  \end{IEEEbiographynophoto}
  
  \begin{IEEEbiographynophoto}{Xiyu Meng}
  is currently a Ph.D. student in the State Key Lab of CAD\&CG, Zhejiang University. He received his B.S. degree in Thermal Engineering from Zhejiang University.
  \end{IEEEbiographynophoto}
  
  \begin{IEEEbiographynophoto}{Mengye Xu} received her B.E. degree in Digital Media Technology from Zhejiang University.
  \end{IEEEbiographynophoto}
  
  \begin{IEEEbiographynophoto}{Yu Liao}
  is currently a graduate student at the Human-Computer Interaction Institute at Carnegie Mellon University. She received her B.E. degree in Digital Media Technology from Zhejiang University.
  \end{IEEEbiographynophoto}
  
  \begin{IEEEbiographynophoto}{Haidong Zhang}
  received the Ph.D. degree in Computer Science from Peking University, China. He is a Principal Architect at Microsoft Research Asia. His research interests include visualization and human-computer interaction.
  \end{IEEEbiographynophoto}
  
  \begin{IEEEbiographynophoto}{Yingcai Wu}
    is a Professor at the State Key Lab of CAD\&CG, Zhejiang University. His main research interests are in information visualization and visual analytics, with focuses on sports science and urban computing. He received his Ph.D. degree in Computer Science from the Hong Kong University of Science and Technology. Prior to his current position, Dr. Wu was a postdoctoral researcher at the University of California, Davis from 2010 to 2012, and a researcher in Microsoft Research Asia from 2012 to 2015. For more information, please visit http://www.ycwu.org.
    \end{IEEEbiographynophoto}


%\vfill


% \newpage
% \input{content/10_appendix}

\end{document}